\documentclass[11pt]{article}
\pdfoutput=1
\usepackage{jheppub}
\usepackage{amssymb,amsmath,amstext,amsfonts}
\usepackage{bbold,ulem,bm}
\usepackage{graphics}
\usepackage{mathtools}
\usepackage{hyperref}
\usepackage{color}

\usepackage{graphicx}
\usepackage{subcaption}
\usepackage{mathrsfs}
\usepackage{multirow} 
\newcommand{\beq}{\begin{equation}}
\newcommand{\eeq}{\end{equation}}
\newcommand{\bal}{\begin{aligned}}
\newcommand{\eal}{\end{aligned}}
\newcommand{\bea}{\begin{eqnarray}}
\newcommand{\eea}{\end{eqnarray}}

\def\a{a}

\def\h{c}
\def\g{f}

\newcommand{\mathsym}[1]{{}}

\renewcommand\({\left(}
\renewcommand\){\right)}
\renewcommand\[{\left[}
\renewcommand\]{\right]}

\newcommand\mpl{M_{\rm Pl}}

\def\be{\begin{equation}}
\def\ee{\end{equation}}
\def\ba{\begin{eqnarray}}
\def\ea{\end{eqnarray}}

\def\M{\mathcal{M}}

\def\H{\mathcal{H}}

\def\({\left(}
\def\){\right)}

\def\Nrel{N_{\mathrm{relax}}}
\def\Trh{T_{\mathrm{RH}}}
\def\hnsr{h_{\rm{ {\bar s} {\bar r} } }}
\def\hcl{h_{\mathrm{cl}}}
\def\Nmin{N_{{\rm \star}}}
\def\hmax{h_{{\rm max}}}
\def\Nm{N_{{\rm m}}}
\def\a{a}

\def\Vinf{V^{{\rm inf}}}
\def\f{f}
\def\h{h}
\def\be{\begin{equation}}
\def\ee{\end{equation}}
\def\ba{\begin{eqnarray}}
\def\ea{\end{eqnarray}}

\title{\centering\huge \bf  Higgs vacuum (in)stability during inflation\bigskip\\ \LARGE The dangerous relevance of de Sitter departure and Planck-suppressed operators}

\author{Jacopo Fumagalli,}
\author{S\'ebastien Renaux-Petel}
\author{and John W. Ronayne}

\emailAdd{fumagall@iap.fr}
\emailAdd{renaux@iap.fr}
\emailAdd{ronayne@iap.fr}

\affiliation{Institut d'Astrophysique de Paris, GReCO, UMR 7095 du CNRS et de Sorbonne Universit\'e, 98bis boulevard Arago, 75014 Paris, France}

\abstract{The measured Standard Model parameters lie in a range such that the Higgs potential, once extrapolated up to high scales, develops a minimum of negative energy density. This has important cosmological implications. In particular, during inflation, quantum fluctuations could have pushed the Higgs field beyond its potential barrier, triggering the formation of anti-de Sitter regions, with fatal consequences for our universe.
	By requiring that this did not happen, one can in principle connect (and constrain) Standard Model parameters with the energy scale of inflation. In this context, we highlight the sensitivity of the fate of our vacuum to seemingly irrelevant physics. In particular, the departure of inflation from an exact de Sitter phase, as well as Planck-suppressed derivative operators, can, already and surprisingly, play a decisive role in (de)stabilizing the Higgs during inflation.
	Furthermore, in the stochastic dynamics, we quantify the impact of the amplitude of the noise differing from the one of a massless field, as well as of going beyond the slow-roll approximation by using a phase-space approach. On a general ground, our analysis shows that relating the period of inflation to precision particle physics requires a knowledge of these ``irrelevant" effects.}

\begin{document}
\vspace*{-1cm}
\maketitle



\newpage
\section{Introduction}

One of the main surprises after the discovery of the Higgs boson \cite{Chatrchyan:2012xdj,Aad:2012tfa} was the fact that the measured values of the Standard Model (SM) parameters lie exactly within the boundary region that separates where the SM would develop, or not, a true vacuum of negative energy density once extrapolated up to the Planck scale (see \cite{Hung:1979dn,Isidori:2001bm,Sher:1993mf,Casas:1996aq,Ellis:2009tp,EliasMiro:2011aa} and \cite{Degrassi:2012ry,Buttazzo:2013uya,Branchina:2013jra} for studies before and after the Higgs discovery). The energy scale at which this instability takes place is extremely sensitive to the boundary conditions measured at the electroweak (EW) scale. In particular, the central values of the measured top and Higgs masses hint that our vacuum is metastable, i.e. it is not the true vacuum but its lifetime is larger than the age of our Universe.\footnote{The addition of Planck-suppressed operators can significantly influence the tunneling rate from the false to the true vacuum \cite{Branchina:2013jra}. However, it has been shown that a small value of the non-minimal coupling is enough to wash out the effect of these higher-order operators \cite{Branchina:2019tyy}.}

As it was phrased in a recent review \cite{Markkanen:2018pdo}: a metastable vacuum, by definition, has implications that can only be studied in the context of the cosmological history. Even if today the lifetime of our vacuum is much larger than the age of our Universe, assuming a period of inflation implies that, in the very early Universe, the Higgs (when it is not the inflaton) behaved as a test scalar field in a (quasi) de Sitter background. There, stochastic kicks could have pushed it beyond the potential barrier towards the true vacuum, until the point of forming anti de Sitter (AdS) regions fatal for our universe \cite{Espinosa:2015qea,East:2016anr}. For good reasons (e.g. you reading this sentence), we have to enforce that no such regions formed in our past light-cone. This request brings interesting implications. 
The shape of the Higgs potential depends on the measured SM parameters, while the size of the stochastic kicks is of  
order of the Hubble rate $H$, which is directed linked to the energy scale of inflation. Thus, the two can be related: given the measured SM parameters one can constrain $H$ \cite{Espinosa:2007qp,Kobakhidze:2013tn,Enqvist:2014bua,Hook:2014uia,Kearney:2015vba,Espinosa:2015qea,East:2016anr,Jain:2019wxo}. Conversely,  assuming a detection of primordial gravitational waves one can constraint the SM parameters \cite{Hook:2014uia, Kearney:2015vba,Franciolini:2018ebs}. Furthermore, after inflation, the oscillations of the inflaton induce tachyonic excitations of the Higgs field that can as well trigger 
a vacuum instability \cite{Herranen:2015ima,Ema:2016kpf,Kohri:2016wof,Enqvist:2016mqj,Postma:2017hbk,Ema:2017loe,Figueroa:2017slm}.

These various studies share two main simplifying assumptions: a constant Hubble scale during inflation, and no new physics between the Standard Model and very high scales.
The latter assumption is motivated by minimality, while the former approximation is motivated by the (typically) slight departure of inflation from a de Sitter phase, and it also has the advantage of making the analysis model-independent.
The purpose of this work is to show that considering departures from a perfect de Sitter background, as well as including Planck-suppressed derivative operators in the analysis, can play a significant role in determining the fate of the Higgs vacuum during inflation.

Consider, as an example, a dimension-six operator of the form 
\begin{equation}\label{operatordes}
\mathcal{O}_{6}=C\frac{2\mathcal{H}^{\dagger}\mathcal{H}}{M^{2}}(\partial\phi)^{2},
\end{equation}
where $\mathcal{H}$ is the Higgs doublet and $\phi$ a generic inflaton.
This operator preserves the would-be (quasi)-shift symmetry of the inflationary sector.
In particular, the request of preserving the flatness of the inflationary potential entails no constraints on $C$ and $M$.
However, higher-order operators like the one in Eq.~\eqref{operatordes}, which are allowed by symmetry and hence compulsory from an effective field theory point of view, generate a non-trivial geometry in the Higgs-inflaton field space manifold; the curvature of this manifold induces an effective mass 
for the Higgs that can stabilize (or further destabilize) it during inflation, similarly to what occurs in the geometrical destabilization of inflation \cite{Renaux-Petel:2015mga}.\footnote{See \cite{Renaux-Petel:2017dia,Garcia-Saenz:2018ifx,Grocholski:2019mot} for studies of the fate of this instability.}
As we will see, the effect is already significant for the conservative choice $M=\mpl$, and it would be completely determinant for $M$ even slightly smaller than $\mpl$.
From Eq.~\eqref{operatordes} it is easy to see that the absolute value of the induced mass is proportional to the first slow-roll parameter, i.e. $-(\partial\phi)^2\simeq\dot{\phi}^2\propto \epsilon= -\dot{H}/H^2$. Hence, our analysis cannot avoid considering an evolving background in which the time dependence of the Hubble parameter $H$ is taken into account. 
More generally, irrespective of the impact of higher-order derivative operators, we will show that the inevitable time dependence of the inflationary background influences in a non-trivial manner both the classical and the stochastic dynamics of the Higgs, and hence its cosmological fate.
Eventually, in the presence of operators that induce an effective mass for the Higgs, 
be they derivative operators like in Eq.~\eqref{operatordes} or non-minimal couplings like in Refs.~\cite{Herranen:2014cua,Espinosa:2015qea}, taking into account the fact that the stochastic noise of a light field differs from the one of an exactly massless field, as it is usually done, has an important impact on the final results.

Summarizing, we study the relevance of the following and previously neglected physics on the Higgs (in)stability during inflation:
\begin{itemize}
	\item \textit{$H$ being not exactly constant.} Inflation takes place in a quasi de Sitter background. Different models will determine different evolutions of $H$ and in turn different stochastic dynamics of the Higgs field. 
	\item \textit{The variance of the noise} deviating from the almost massless case. Parameterizing the random kicks with  $H/2\pi$ is accurate only when there is a large hierarchy between the mass of the field and $H$.
	\item \textit{Planck-suppressed derivative operators.} Couplings like the one in Eq.~\eqref{operatordes}, as well as more general ones that respect the shift symmetry of the inflationary sector, modify the effective mass of the Higgs, for instance by inducing a non-trivial geometry in the Higgs-inflaton target space.
	\item \textit{Considering a stochastic approach in phase space.} Deviations of inflation from a strict slow-roll phase is communicated to the spectator Higgs, notably at the end of inflation. We therefore take into account stochastic effects beyond the slow-roll regime.
\end{itemize}
While the improvement coming from considering the stochastic approach in phase space has
a minor impact on the final results, we show explicitly that the first three effects in general play a crucial role in determining the fate of the instability. As a result, for fixed boundary conditions, i.e. measured Standard Model parameters and scale of inflation $H$, we obtain, for 
different inflationary models and Planck-suppressed operators, outcomes for the fate of the instability that are different, sometimes by orders of magnitude, from the benchmark analysis in which the aforementioned effects are neglected.

Moreover, the time dependence of the background and of the various effects contributing to the Higgs dynamics prompts us to introduce important conceptual novelties in the analysis. A non-static effective potential combined with the stochastic diffusion of the Higgs leads to a new procedure (explained in Sec. \ref{matching}) to compute the fraction of Hubble patches in AdS in our past light cone. 
In particular, when matching the stochastic and classical dynamics, we pay attention to the fact that patches already in AdS cannot be rescued, together with the finite time it takes for them to form when the Higgs backreaction cannot be neglected anymore.

Our main numerical results are displayed in figures \ref{survival}-\ref{effect-b1-survival} and \ref{mtt}-\ref{mhh}. In the first two, we show the impact of our analysis in shaping the constraints on the Hubble scale, for two different background evolutions, each with and without derivative Planck-suppressed operators. In short, the effects we studied, that might be considered
negligible, are instead often crucial to correctly estimate the bounds on $H$.  From a different perspective, in the latter two figures, we also present 
our results by looking at how a given Hubble scale constrains the top and Higgs masses within their experimental error bars.
Remarkably, there exists a degeneracy between values of $H$ separated by different orders of magnitude on one side, and effects coming from the time-dependent background or derivative Planck-suppressed operators on the other side. Therefore, even with the assumption of the SM being valid up to the Planck scale, it appears unlikely that a future detection of primordial gravitational waves would, on its own, enable us to constrain SM parameters like the top mass.

The paper is organized as follows. In Sec. \ref{Planck1} we describe the various effects that contribute to the classical dynamics of the Higgs during inflation. In particular, we highlight the effects of derivative Planck-suppressed operators. We introduce stochastic effects and the probability distribution function (PDF) of the Higgs in Sec. \ref{sec:valid}, before studying each of the effects listed above in Sec. \ref{secevo}.
In Sec. \ref{anthropic} we define the criteria required to avoid patches of AdS in our past light cone, while in Sec. \ref{matching} we outline the procedure used to estimate the amount of these patches in a time-dependent setup, and give analytical estimates in Sec. \ref{secmod1}.
The last part \ref{secres} is devoted to our (numerical) results. There we show explicitly the sensitivity of the fate of the Higgs instability 
to the different effects studied in this work.  Conclusion and outlooks are provided in Sec. \ref{seccon}.

\subsubsection*{Falling in the AdS vacuum (or a brief story of an AdS patch)}
\begin{figure}
	\centering
	\includegraphics[width=0.75\linewidth]{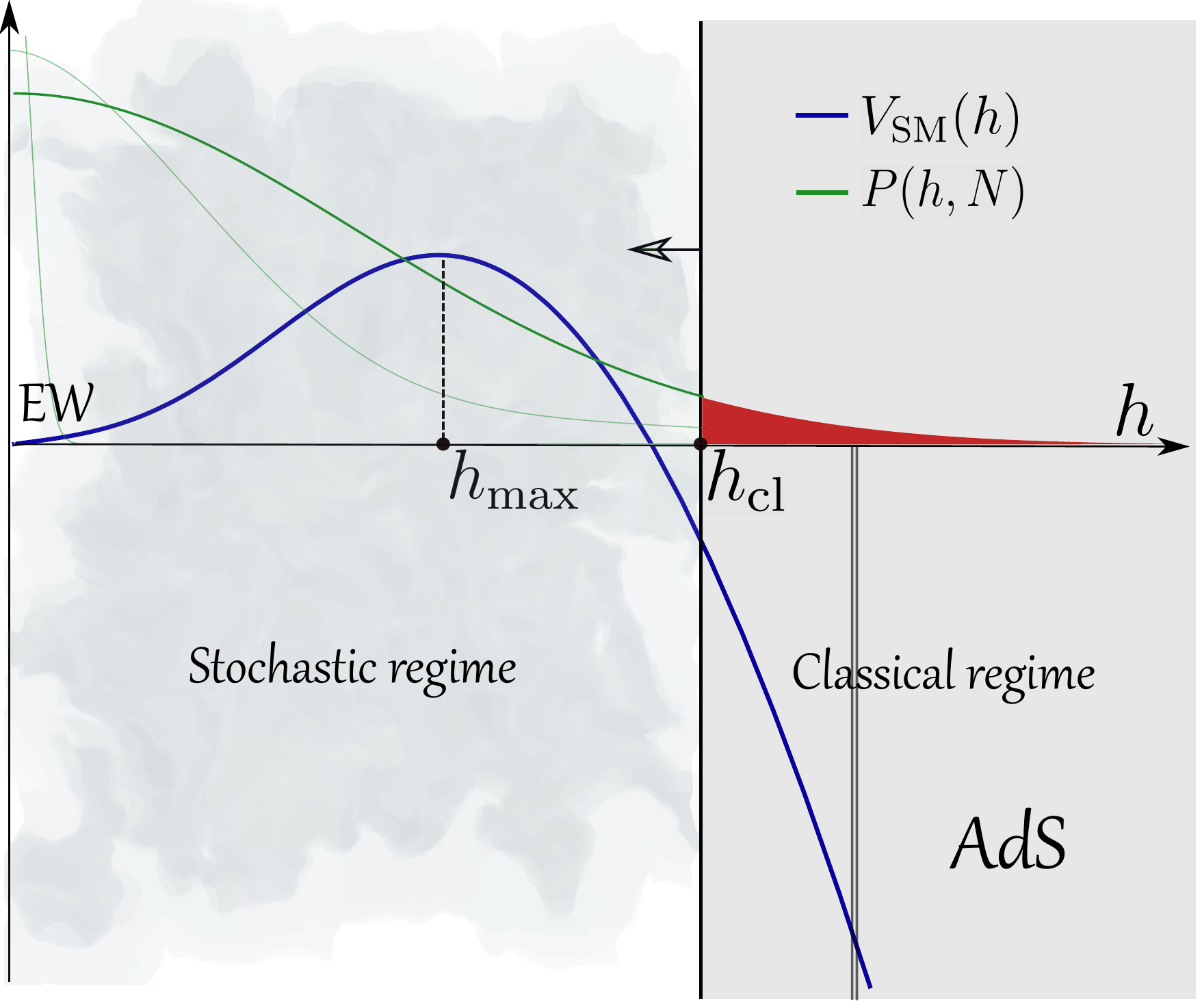}
	\caption{\textit{Dynamics of the Higgs field during inflation. In blue: sketch of the SM potential which, for the current best measurement of the SM parameters, develops an instability at large field values. We denote the position of the potential barrier by $h_{\mathrm{max}}$, and the Higgs value beyond which the classical dynamics dominates over the quantum jumps by $\hcl$ (Eq. \eqref{had}). This ``point of classicality" evolves in a time-dependent background. In green the PDF (at three different times) that gives the probability of finding a given Hubble patch with Higgs value $h$ after $N$ $e$-folds of inflation. We trust the PDF up to $\hcl$, for larger values we evolve the tail classically to compute the fraction of Hubble patches that are in AdS at the end of inflation, see Sec. \ref{matching}.}}
	\label{falling}
\end{figure}
The fate of a spacetime region in which a scalar field is falling towards a negative energy vacuum, as well as the fate of an AdS patch embedded
in a de Sitter geometry, represent non-trivial general relativity problems. As a matter of fact, these issues were not well understood until very recently \cite{Espinosa:2015qea,East:2016anr}. 
Thus, before starting, we briefly brush over the current understanding of these phenomena.
As mentioned, for central values of the SM parameters, the effective quartic coupling of the Higgs potential becomes negative,
hence developing an instability, at field values much less than the Planck scale.
We call $\hmax$ the value of the Higgs field at which the SM potential has its maximum. 
For large enough field values, the classical motion, led by the SM potential, dominates
over the stochastic quantum contributions.
From this point on, that we label as $h_{\rm cl}$ (defined in Eq. \eqref{had}), the Higgs starts to roll down classically towards the true vacuum. 
Initially, when the Higgs experiences the negative part of its potential, the total energy density is still completely dominated by the positive contribution coming from the inflationary background $V(\phi)\simeq 3\mpl^2 H^2$. Once the energy density of the Higgs sector becomes large enough, it
strongly backreacts on spacetime, to the point of eventually generating an anti-de Sitter (AdS) patch.
In this respect, the results first obtained analytically in \cite{Espinosa:2015qea} were later substantially confirmed with 
full GR simulations in \cite{East:2016anr}: in Hubble patches in which the Higgs field exceeds $h_{\rm cl}$, after a finite time inflation terminates locally, leading to a crunching region surrounded by a causally disconnected one of negative energy density, i.e. an AdS patch.
The latter persists during inflation and, as a first approximation, expands comovingly with the ambient de Sitter spacetime. Conversely, after inflation (in an approximately Minkowski background), the AdS patch
expands at the speed of light, engulfing the surrounding space-time which therefore cannot be in the electroweak vacuum today. This is the reason to demand that there was not a single AdS Hubble patch in our past light cone, see Eqs.~\eqref{prob}-\eqref{prob2}. Furthermore, since the 
generation and evolution of an AdS bubble is not fatal to the ambient inflationary spacetime, it is possible to count the fraction of volume transmuting into AdS by using a probability density function (PDF) following a Fokker-Planck equation, augmented with necessary precautions listed
in Sec. \ref{section3}.


\section{The stochastic Higgs during inflation}
\label{stochastic}

\subsection{Classical dynamics and effect of derivative operators }\label{Planck1}

Here we describe the various effects that we take into account and that determine the classical dynamics of the Higgs during inflation, postponing to the next section the inclusion of stochastic effects. The radial SM Higgs $h$ is taken as a spectator field with (initially) no role in the inflationary dynamics, the latter being driven by an inflaton field $\phi$ endowed with a potential $V(\phi)$. For simplicity, the other operators that we consider are taken to respect the shift symmetry $\phi \to \phi+{\rm const.}$, hence we do not include Higgs-inflaton couplings in the potential.\footnote{For instance, operators like
	\begin{equation}\label{spoil}
	\mathcal{O}_{n+2}=C_{n+2} \frac{\phi^{n}h^2}{\mpl^{n-2}} 
	\end{equation}
	induce an effective mass for the Higgs (see for instance \cite{Fairbairn:2014zia,Kamada:2014ufa}). However, contrary to shift-symmetric derivative operators that we study, these operators are tightly constrained
	by the request of not spoiling at loop level the flatness of the inflationary potential. In fact, even if the operator \eqref{spoil} does not influence the dynamics at tree level for $h=0$ on the background, it generates a $\phi$ dependent one-loop contribution to the potential. Thus, all the Wilson coefficients in \eqref{spoil} have to be rather small, with bounds depending on the particular form of the inflationary potential. Reasoning as in \cite{Lebedev:2012sy}, one can easily show that for $V(\phi)=1/2 \,m^2 \phi^2$ for instance, one has $C_{n+2}\lesssim 10^{-(4+n)}$.} However, we take into account the non-minimal coupling of the Higgs to the spacetime curvature $R$, and two-derivative higher-order operators. The resulting total Lagrangian reads
\begin{equation}\label{startingaction}
\mathcal{L}=-\frac{1}{2}G_{IJ}\partial\varphi^J\partial\varphi^I-V(h)-\Vinf(\phi),\qquad \varphi^I=\{\phi,h\},
\end{equation}
where the higher-order operators are included in the inflaton-Higgs field space metric 
$G_{IJ}$, to which we come back below, and $V(h)$ is defined as
\begin{equation}\label{higgspot}
V(h)=V_{\rm{SM}}-\frac{\xi h^{2}}{2}R,\quad V_{\rm{SM}}= \frac{\lambda_{\rm{eff}} (\mu(h))}{4}h^4\,.
\end{equation}
Here, $V_{\rm{SM}}$ is the renormalization group (RG) improved Standard Model Higgs potential. In our analysis, it has been computed at NNLO \cite{Degrassi:2012ry}, using the two-loop effective potential \cite{Ford:1992pn}, two-loop matching conditions at the EW scale \cite{Buttazzo:2013uya} and three-loop beta-functions \cite{Luo:2002ey,Mihaila:2012fm,Chetyrkin:2012rz}. 
$\lambda_{\rm {eff}}$ is the two-loop effective quartic coupling in the Landau gauge defined in Ref.~\cite{Buttazzo:2013uya}, with the contribution from the anomalous dimension already absorbed in a field redefinition of $h$, which facilitates the analysis and  significantly reduces the gauge dependence of the potential \cite{Espinosa:2015qea}. The optimal choice for the renormalization scale has been chosen to take into account the (quasi) de Sitter background, i.e. to keep higher-order terms under control we use $\mu^2 \simeq h^2 +12 H^2$ \cite{Markkanen:2018bfx}, where sub-leading slow-roll corrections are neglected, although it has been shown that considering other linear combinations such as $\mu^2=\alpha h^2+\beta H^2$ has negligible impact \cite{Herranen:2014cua}. 
We consider generic values of the non-minimal coupling as $\xi=0$ is not a fixed point of the renormalization group flow, i.e. if it is set to zero at one scale it will be different from zero at any other scale. Note that in our convention, a negative $\xi$ tends to stabilize the Higgs.\footnote{We work with signature $(-+++)$, such that $R=(12H^2+6\dot{H})$. 
We sometimes compare our results to the ones of Refs. \cite{Espinosa:2015qea,Franciolini:2018ebs}, which
use signature $(+-\,-\,-)$ such that $R=-(12H^2+6\dot{H})$. However negative values of $\xi$ still correspond to a stabilizing effect since we use a sign flip in the definition of the non-minimal coupling in \eqref{higgspot}.}

Before discussing the kinetic terms, it is instructive to use a multifield point of view, in which the Higgs direction is identified with the entropic direction in a two-field model (albeit a special one in which the Higgs is not contributing to the background dynamics). In this context, its (super-Hubble) fluctuations are known to acquire the effective mass (see e.g. \cite{Sasaki:1995aw,Bartjan-linear})\footnote{This multifield formalism is usually formulated in the Einstein frame, but relative corrections compared to the Jordan frame used in this paper are in $h^2/\mpl^2$, so they are completely negligible. This is the implicit point of view we also use when incorporating the non-minimal coupling in the Higgs potential \eqref{higgspot}.}
\begin{equation}
V_{;\,ss}+3 H^2 \eta_\perp^2+\epsilon R_{\rm{fs}}H^2 \mpl^2\,,
\label{ms2}
\end{equation}  
where semicolons stay for covariant derivatives with respect to the field space metric $G_{IJ}$, the subscript $s$ indicates a projection along the entropic direction, $\eta_\perp$ is a dimensionless parameter that measures the deviation of the trajectory from a field space geodesic, and $R_{\rm{fs}}$ denotes the curvature of the field space. This formula makes it clear that the effective mass of the Higgs during inflation is not simply given by $V^{''}(h)=V^{''}_{\rm{SM}}- \xi R$. In particular, non-standard kinetic terms contribute to it in general, for instance through the curvature of the field space, but also through non-trivial Christoffel symbols in the covariant derivative $V_{;s s} \simeq V_{;hh}$. In what follows, we confirm these expectations in a simple EFT parameterization of the kinetic terms.

As mentioned above, we consider kinetic terms that respect the (approximate) shift-symmetry of the inflaton, i.e. with $G_{IJ}$ independent of the inflaton. Taking into account the $SU(2)$ gauge symmetry of the electroweak sector further restricts the allowed kinetic terms. Momentarily using the Higgs doublet $\mathcal{H}$, they are of the form 
\begin{equation}
{\cal L}_{\rm kin}=-D^{\mu} \H^\dagger D_{\mu} \H-\frac12\a(\H^{\dagger} \H)(\nabla_{\mu} \phi)^2  - b(\H^{\dagger} \H) \left( \H^\dagger D^{\mu} \H \nabla_\mu \phi+{\rm h.c.} \right)
\label{kinetic-term-doublet}
\end{equation}
where $D_\mu$ denotes the gauge derivative, $a$ and $b$ are generic functions, and the kinetic term involving only the Higgs can always be put in a canonical form. As usual, we parameterize the effect of high-scale physics by expanding $a$ and $b$ in powers of $\H^{\dagger} \H/M^2$, where $M$ denotes the cutoff of the theory.
In terms of the radial Higgs field, we thus write
\begin{equation}
{\cal L}_{\rm kin}=-\frac12 (\partial h)^2-\frac12 \left(1-2C_6 \left(\frac{h}{M} \right)^2+\ldots \right)  (\partial \phi)^2 - C_5 \frac{h}{M} \partial h \partial \phi  \left(1+{\cal O} \left(\frac{h}{M} \right)^2+\ldots \right)\,,
\label{kinetic-term}
\end{equation}
where all coefficients are thought to be of order one, and the terms in $C_5$ and $C_6$ correspond respectively to dimension $5$ and $6$ operators. Other higher-order operators, corresponding to higher powers of $h/M$, can be kept, but have negligible impact for our purposes, i.e. we only keep dangerous irrelevant operators. Eventually, note that  the field space defined by the kinetic terms \eqref{kinetic-term} is curved for generic values of the parameters $C_5$ and $C_6$.\\[-0.3cm]

The classical background equations of motion deduced from \eqref{startingaction} read $\ddot \phi^I +\Gamma^I_{JK} \dot \phi^J \dot \phi^K+3H \dot \phi^I+G^{IJ}V_{,J}=0$, where $\Gamma^I_{JK}$ denotes the Christoffel symbols of the metric $G_{IJ}$, i.e.
\ba
\ddot \phi-2C_5 C_6 \frac{h^2}{M^3}  \dot \phi^2-4 C_6 \frac{h}{M^2} \dot \phi \dot h+\frac{C_5}{M} \dot h^2 +3 H \dot \phi+\Vinf_{,\phi}-C_5 \frac{h}{M} V_{,h}\simeq 0\,,\label{eom-phi} \\
\ddot h+2 C_6 \frac{h}{M^2} \dot \phi^2+4 C_5 C_6 \frac{h^2}{M^3}  \dot \phi \dot h-C_5^2 \frac{h}{M^2} \dot h^2+3H \dot h+V_{,h} -C_5 \frac{h}{M} \Vinf_{,\phi} \simeq 0\,, \label{eom-h}
\ea
where we kept for each term only its dominant part in $h/M$.
Using these equations, one can easily show the self-consistency of the regime where the Higgs is considered as a spectator field, with no backreaction on the inflaton, and where the dynamics obeys:
\ba
3 H \dot \phi+\Vinf_{,\phi} \simeq 0\,,\label{eom-phi-approximate} \\
\ddot h  +3H \dot h+V_{,h} + \left(-C_5 \frac{\Vinf_{,\phi}}{M}+2 C_6 \frac{\dot \phi^2}{M^2}  \right) h \simeq 0\,, \label{eom-h-approximate} \\
3 H^2 \mpl^2 \simeq \Vinf(\phi)+\frac12 \dot \phi^2\,. \label{H-approximate}
\ea
For this, note that the typical velocity of the Higgs is $\dot h \sim H^2$ (as the analysis below will confirm), so that the kinetic energy is completely dominated by the inflaton: $\dot h^2/\dot \phi^2 \sim H^2/(\epsilon \mpl^2) \sim {\cal P}_{\zeta} \sim 10^{-10}$,
where ${\cal P}_{\zeta}$ denotes the amplitude of the primordial curvature power spectrum, and one consistently used the fact that $\epsilon$ is directly related to the velocity of the inflaton:
\begin{equation}
\epsilon \equiv -\frac{\dot H}{H^2} \simeq \frac{\dot{\phi}^2}{2 H^2 \mpl^2}\,.
\label{epsilon}
\end{equation}
All terms neglected in \eqref{eom-phi-approximate}-\eqref{epsilon} are hence suppressed compared to leading-order ones by (combination of) powers of $h/M$, $h/\mpl$, $H/M$ and ${\cal P}_\zeta$. \\[-0.3cm]

Summarizing: derivative operators in \eqref{kinetic-term} have a negligible impact on the inflaton, but they modify the dynamics of the Higgs, whose evolution \eqref{eom-h-approximate} can be intuitively understood as the one of a canonical field subject to the time-dependent effective potential
\begin{equation}\label{effpot}
V_{\rm{eff}}=V_{\rm{SM}}(h)+\frac{H^2 h^2}{2} \left(-12 \xi \left( 1-\frac{\epsilon}{2}\right)-3 C_5\, {\rm sign}(\Vinf_{,\phi})\, \sqrt{2\epsilon} \frac{\mpl}{M}+ 4 C_6 \epsilon \frac{\mpl^2}{M^2} \right) \,.
\end{equation}
Here, we used Eqs.~\eqref{eom-phi-approximate} and \eqref{epsilon} to express the effective potential in terms of  the first slow-roll parameter $\epsilon$. 
The three terms in parenthesis correspond to the effects of the non-minimal coupling, and of the dimension $5$ and $6$ derivative operators. Each generate quadratic contributions to the potential, whose effective mass in Hubble units are set respectively by $\xi$, $\sqrt{\epsilon}\mpl/M$ and $\epsilon\mpl^2/M^2 $. Hence, although the effects of the derivative operators may seem innocuous at first sight, as they are slow-roll suppressed, a second thought reveals that they can play an important role in the dynamics of the Higgs field, in the same way as a small value of $\xi$ can modify the fate of the Higgs instability during inflation \cite{Herranen:2014cua,Espinosa:2015qea}. Furthermore, it is important to stress that, motivated by minimality, we will focus on operators suppressed by the Planck scale, i.e. $M=\mpl$, but effects are obviously even more important if one considers values of $M$ even slightly smaller than $\mpl$.
Eventually, let us note that the effects of the derivative operators are tied to the non-zero value of $\epsilon$, or equivalently to the slight breaking of the inflationary shift symmetry by the potential. In general, one expects this breaking to be communicated through loops to the kinetic sector, i.e. one expects derivative couplings that also slightly break the shift symmetry. We leave the study of such a general setup to future works, and content ourselves with assessing the impact of the operators in \eqref{kinetic-term}.\\[-0.3cm]

Following the interpretation of Eq.~\eqref{effpot} as the effective potential governing the dynamics of the spectator Higgs field, it is natural to define its effective mass as
\begin{equation}
\M^2 \equiv \frac{\partial^2V_{\rm{eff}}}{\partial h^2} =V^{''}_{\rm{SM}}+H^2 \left(-12 \xi \left( 1-\frac{\epsilon}{2}\right)-3 C_5 \sqrt{2 \epsilon} \frac{\mpl}{M}+ 4 C_6 \epsilon \frac{\mpl^2}{M^2} \right)\,,
\label{M2}
\end{equation}
where we chose ${\rm sign}(\Vinf_{,\phi})=1$, which one can always do for monotonous potentials by redefining $\phi \to -\phi$. It is instructive to compare this to the effective mass \eqref{ms2}. Under the same approximations as above, one can easily show that the two expressions coincide at leading-order in $h/M$, with a negligible contribution from the bending (the second term in \eqref{ms2}), $V_{;ss} \simeq V_{,hh}-C_5 \Vinf_{,\phi}/M$ reproducing the first three terms in \eqref{M2}, and $\epsilon R_{\rm{fs}} \mpl^2 \simeq 4 C_6 \epsilon \mpl^2/M^2$ giving rise to the last term. While the effect of the dimension-six operator can thus be explained by its contribution to the field space curvature, as mentioned in the introduction, the effect of the dimension-five operator comes from its contribution to the covariant Hessian of the potential.

\subsection{Stochastic dynamics}\label{secsto}
\label{sec:valid}

As inflation proceeds, initially sub-Hubble fluctuations of the Higgs field exit the Hubble radius and feed its infrared dynamics \cite{Starobinsky:1986fx,Starobinsky:1994bd}.
This stochastic evolution is usually modeled by the simple Langevin equation
\begin{equation}
\label{langevin} 
\frac{d h}{dN}+\frac{1}{3H^2}\frac{\partial V_{\rm{eff}}}{\partial h}=\eta(N)\,,
\end{equation}
where here and in the remainder of this paper, $h$ denotes the super-Hubble coarse-grained part of the Higgs field. $N$ is the number of $e$-folds of inflation, and $\eta$ is a Gaussian white noise with variance the power spectrum of the Higgs fluctuations when they join the IR sector (more about the factor $\f$ below in section \ref{section-f}):
\begin{equation}
\label{noise}
\langle \eta(N)\eta(N')\rangle =\(\frac{H \f}{2\pi}\)^2 \,\delta(N-N')\,.
\end{equation} 
Stochastic effects have received a renewed attention in the past years (see e.g. \cite{Fujita:2014tja,Burgess:2014eoa,Vennin:2015hra,Burgess:2015ajz,Vennin:2016wnk,Moss:2016uix,Hardwick:2017fjo,Grain:2017dqa,Collins:2017haz,Prokopec:2017vxx,Tokuda:2017fdh,Hardwick:2018sck,Tokuda:2018eqs,Pinol:2018euk,Hardwick:2019uex,Markkanen:2019kpv}).
However, despite substantial progress, a general theory quantifying the theoretical errors of Eqs.~\eqref{langevin}-\eqref{noise} is still lacking, concerning for instance the approximations of a Markovian dynamics or the Gaussianity of the noise. Given the scope of this paper, we will very conservatively use Eqs.~\eqref{langevin}-\eqref{noise} (and a phase-space generalization in section \ref{fullstoca}). Below we discuss in detail their practical implementation and consequences, but for the moment, it is enough to mention their main characteristics.

In particular, long-wavelength fluctuations are substantially generated, corresponding to $\f \simeq 1$ in \eqref{noise}, only if the mass of the scalar field fluctuations is light enough, i.e. with $\M^2$ in \eqref{M2} verifying $\M^2 \ll H^2$, whereas fluctuations are exponentially suppressed if $\M^2>9/4 H^2$.
In the relevant range of values of $h$ that we will be led to consider, the effect of the SM potential is negligible, as we will discuss in more detail in the next section. Considering for the moment only the non-minimal coupling, like in the current literature, $\M^2 \simeq -12 \xi H^2$ and several cases have to be distinguished.
For $\xi<-3/16\equiv\xi_1$ the Higgs fluctuations are suppressed and there are no stochastic kicks. Thus, if the Higgs starts below the instability scale, i.e. $|h|\lesssim \hmax$, a non-minimal coupling $\xi<\xi_1$ is enough to ensure stability during inflation \cite{Herranen:2014cua}. However, $\xi$ additionally has to obey $\xi \gtrsim -{\cal O}(1)$ to ensure stability during (p)reheating \cite{Herranen:2015ima,Kohri:2016wof,Figueroa:2017slm}.\footnote{During preheating the Ricci scalar rapidly oscillates about zero. When the induced mass term is negative, the associated tachyonic instability can lead to efficient particle production triggering the vacuum instability. The higher-order operators considered in this work could potentially have a similar effect, and a careful study of the preheating phase might constraint the size of the coefficients $C_5,C_6$. However, since $\mathcal{O}_6\propto C_6 \dot{\phi}^2$ and $\dot{\phi}^2$ is always positive, we can already argue that, for $C_6>0$, no constraint would arise from this effect.} In order to check if values of $\xi > \xi_1$ are compatible with our universe, one has to take into account stochastic effects.\footnote{For $0<\xi<{\cal O}(1)$, the effective potential acquires another minimum, but the Higgs is still light, so that the stochastic approach is valid, see e.g. \cite{Franciolini:2018ebs}.}
The mass terms generated by the higher-order operators that we consider in this paper are proportional to $\sqrt{\epsilon}$ or $\epsilon$ and hence are negligible at the beginning of inflation, at least for $M \simeq \mpl$. Thus, if the value of $\xi$ is such that stochastic effects are inefficient, then our terms will not change this drastically. Their inclusion may modify the upper bound derived from studying the post-inflationary evolution but this is beyond the scope of this work. However, for a given value of $\xi$ for which stochastic effects are important, the fate of the Higgs does depend on the higher-order operators, which become increasingly important as inflation proceeds and $\epsilon$ grows.

From the Langevin equation \eqref{langevin}-\eqref{noise}, one can write the Fokker-Planck (FP) equation for the probability distribution function (PDF) $P(h,N)$ that gives the probability (given some initial conditions) that in a particular Hubble volume the Higgs acquires the value $h$ after $N$ $e$-folds of inflation\footnote{It is worth stressing that even if the instability scale $\hmax$ or other intermediate quantities are gauge dependent quantities, the probabilities derived from the FP equation are not \cite{Espinosa:2015qea}.}:
\begin{equation}
\frac{\partial P}{\partial N}=\frac{\partial}{\partial h}\left(\frac{\partial V_{\rm{eff}} /\partial h}{3H^2}P\right)+\frac{\partial^2}{\partial h^2}\left(\frac{H^2}{8\pi^2} \f^2 P\right)\,.
\label{Fokker}
\end{equation}
We label any finite integral of the PDF with the notation 
\begin{equation}
\mathcal{P}(|h|<\Lambda,N)\equiv\int_{-\Lambda}^{\Lambda}P(h',N)dh' \,.
\label{P-integrated}
\end{equation}
Given that our initial conditions at $N=0$ will consist of the Hubble patch that is the progenitor of our observable universe, $\mathcal{P}(|h|<\Lambda,N)$ can be equivalently interpreted as the fraction of corresponding volume at time $N$ in which $|h|<\Lambda$. The next section is dedicated to the study of the evolution of $P(h,N)$, which constitutes the building block of our analysis, and can be used in other contexts. However, owing to the backreaction of the Higgs on spacetime when the former falls towards the true vacuum, it is worth stressing 
at this stage that the study of the cosmological fate of the Higgs requires additional efforts beyond the computation of the PDF, which will be the focus of section \ref{section3}.

\subsection{Evolution of the variance}
\label{secevo}

\subsubsection{Gaussian approximation}
\label{Gaussian-PDF}

A fact that considerably simplifies the stochastic analysis is that the contribution from $V_{\rm{SM}}$ to the effective potential \eqref{effpot} is negligible in the regime where stochastic effects play an important role. This is obviously not true anymore for large values of $h$ such that the potential is steep and the Higgs classically fall towards the AdS vacuum, which is the object of Sec. \ref{section3}. Neglecting the running of $\lambda$ for the sake of the argument, the ratio between the SM contribution to the mass term and the $\xi$ one is $(\lambda/24 \xi) (h/H)^2$, and the ratio between the SM contribution to the drift and the amplitude of the noise is $(2 \pi \lambda/3) (h/H)^3$. As stochastic effects lead to values of $h$ of order $H$,\footnote{More quantitatively, we will see that, neglecting the SM contribution, typical values of $h^2/H^2$ are of order $3 H^2/(8 \pi^2 \M^2)$ (see Eq.~\eqref{equilibrium}), so that the above first ratio is of order $10^{-4} \lambda/\xi^2$, so indeed well negligible.} these ratios are of order $\lambda \lesssim 10^{-2}$.
We will thus be able to neglect the first term in the mass \eqref{M2}, and most importantly, the SM contribution to the drift term in the Langevin equation \eqref{langevin}, which is thus linear.
Hence, assuming Gaussian initial conditions, the PDF remains Gaussian. This is indeed the case in what follows, as we take as initial conditions for the Higgs values a Dirac delta centered in zero, so that the PDF is centered and only described by its variance. This choice is the one often made in the literature and can be thought of as ``the most optimistic approach", with initial conditions taken $\Nmin$ $e$-folds before the end of inflation, when the largest scales observed today exited the Hubble radius (this number depends on the reheating history, but for definiteness, we conservatively use $\Nmin=60$ in numerical applications).

Denoting the variance by $\sigma^2 \equiv \langle h^2\rangle$, one deduces from the FP equation \eqref{Fokker} that it evolves as
\begin{equation}\label{variance2}
\frac{d \sigma^2(N)}{d N}=-\frac{2\M^2}{3H^2}\sigma^2+\frac{H^2 f^2}{4\pi^2},
\end{equation}
whose solution with initial condition $\sigma(0)=0$ is given by
\begin{equation}
\sigma^2(N)=\frac{1}{4\pi^2}\int^N_0 dN'H^2(N')f^2(N')\exp\(-\frac{2}{3}\int^{N}_{N'} dN''\frac{\M^2(N'')}{H^2(N'')}\),
\label{general-solution}
\end{equation}
where we remind that we label by $N=0$ the time at which the cosmological pivot scale exits the Hubble radius, with $H(N=0)\equiv H_\star$. To better appreciate the effects that we study in this paper, let us first consider the benchmark solution of Eq.~\eqref{general-solution} under the simplifying assumptions that $H$ is constant and $f=1$, i.e. a pure de Sitter phase and stochastic kicks of an exactly massless field. With $\epsilon=0$, the mass term \eqref{M2} simplifies to $\M^2= -12\xi H^2=\mathrm{const}$, and the solution \eqref{general-solution} becomes
\begin{equation}
\sigma^2 = \frac{3 H^4}{8 \pi^2 \M^2}\[1-\exp\(-\frac{2\M^2}{3H^2}N\)\].
\end{equation}
In particular, for the interesting situation of a positive mass term $\xi<0$, the distribution relaxes towards the steady ``de Sitter equilibrium" in a typical time-scale given by $\Nrel \simeq H^2/\M^2=-1/12 \xi$. Thus, for $N \gtrsim\Nrel$, the variance reaches a constant value given by 
\begin{equation}
\label{equilibrium}
\sigma^2_{\mathrm{eq}}=\frac{3 H^4}{8\pi^2\M^2}.
\end{equation}
Obviously, this can occur within the last $\Nmin$ \textit{e}-folds of inflation that we consider only if $\Nrel \lesssim \Nmin$, i.e. if  $| \xi | \gtrsim10^{-3}$ for $\Nmin \simeq 60$. For somewhat smaller values, one can formally consider the limit $\xi \rightarrow 0$ ($\Nrel \rightarrow\infty$), in which case the Higgs simply undergoes free diffusion, with a variance linearly growing with time: $\lim_{\xi\rightarrow 0}\sigma^2=(H^2/4\pi^2) N$.\footnote{Ref.~\cite{Jain:2019gsq} takes into account the possibility of reaching a static distribution in this massless case due to the effects of boundary conditions.}

In the following subsections, we discuss one by one the different effects that make our results differ from the benchmark one \eqref{equilibrium}.

\subsubsection{Deviation from massless noise}
\label{section-f}

When taking into account stochastic effects, a split should be performed between the infrared scales described by the theory, which are sufficient larger than the Hubble radius, and the ultraviolet modes. Incorporating this splitting via a smooth window function is physically motivated but results in a colored noise, which render the analysis technically more involved, and with results hardly depending on details of the window function as it becomes sharp (see e.g. \cite{Winitzki:1999ve,Matarrese:2003ye,Liguori:2004fa}). As a result, a hard cutoff is usually used, with the introduction of a small parameter $w$ such that only modes with $k \leq w a H$ are described by the stochastic theory. We follow this procedure, and for the amplitude of the noise, we use the analytical approximation of the power spectrum of a test scalar field of mass parameter $\M^2$ in de Sitter space, giving rise to the noise power spectrum \eqref{noise} with
\begin{equation}
\label{f-explicit}
f=\left\{
\begin{array}{ll}
\sqrt{\frac{\pi}{2}} w^{3/2} \Big|H^{(1)}_{\nu}(w)\Big|\,, \quad \quad \quad \,\, \nu=\sqrt{9/4-\M^2/H^2} \quad {\rm for} \quad \M^2/H^2 \leq 9/4\\
\sqrt{\frac{\pi}{2}} w^{3/2} e^{-\mu \frac{\pi}{2}} \Big|H^{(1)}_{i \mu}(w)\Big|\,, \quad \mu=\sqrt{\M^2/H^2-9/4} \quad {\rm for} \quad \M^2/H^2 \geq 9/4\,,&
\end{array}\right.
\end{equation}
where $H^{(1)}_{\nu}$ is the Hankel function of the first kind. 
\begin{figure}[t]
	\centering
	\includegraphics[width=0.6\textwidth]{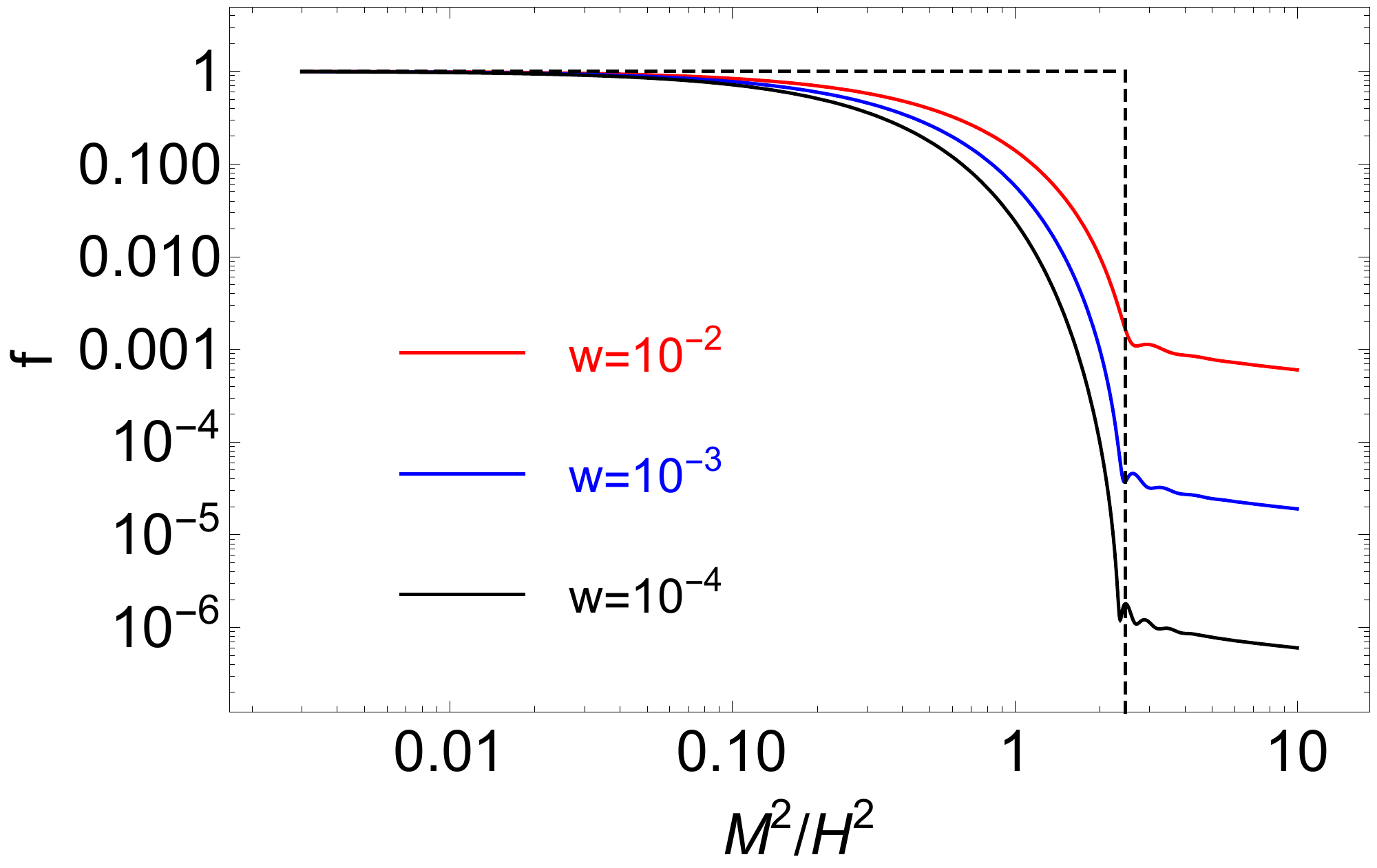}
	\caption{\textit{Dependence on $\M^2/H^2$ of $f$ in Eq.~\eqref{f-explicit}, governing the amplitude of the stochastic noise in \eqref{noise}, for different values of $w$. The dashed line represents the zeroth-order description that is often used, with $f=1$ for $\M^2/H^2<9/4$ and $f$ vanishing for larger values.}}
\label{fig-noise}
\end{figure}
The dependence of $f$ on $\M^2/H^2$ is displayed in figure~\ref{fig-noise} for three different values of $w$. For light enough scalar fields with $\M^2/H^2 \lesssim 10^{-2}$, one recovers the standard amplitude of the noise usually considered in the stochastic formalism, i.e. $f \simeq 1$, with only a percent level deviation for all values of $w$. Naturally, the almost independence on $w$ comes from the fact that such almost massless fields acquire an almost constant amplitude on super-Hubble scales. For $\M^2/h^2 > 9/4$, fluctuations decay rapidly on super-Hubble scales, hence the strong dependence on $w$, and the very small value of $f \lesssim 10^{-3}$, which is well consistent with the zeroth-order description in which such massive fields are considered not to give rise to stochastic fluctuations. The intermediate regime $0.1 \lesssim \M^2/H^2 \lesssim 1$ is more subtle, as stochastic effects can not be neglected then, but the precise value of $f$ depends on the arbitrary choice of $w$. This limitation of the current formulation of the stochastic formalism motivates further studies, which are however beyond the scope of this work. In the rest, we simply use $w=10^{-2}$, noting that our procedure has the advantage of not overestimating the noise in these ``quasi-massive'' situations compared to the zeroth-order description often used, represented in figure~\ref{fig-noise} by the dashed line, 
For instance, in the same de Sitter approximation as in the previous section \ref{Gaussian-PDF}, the equilibrium result \eqref{equilibrium} for the variance is multiplied by $\f^2$, which, for values of $\xi$ as small as $0.01$, already gives rise to a decrease by a factor of $2$.

\subsubsection{Time dependence of $H$}\label{Htimedependence}
\begin{figure}[t]
	\centering
	\includegraphics[width=0.6\textwidth]{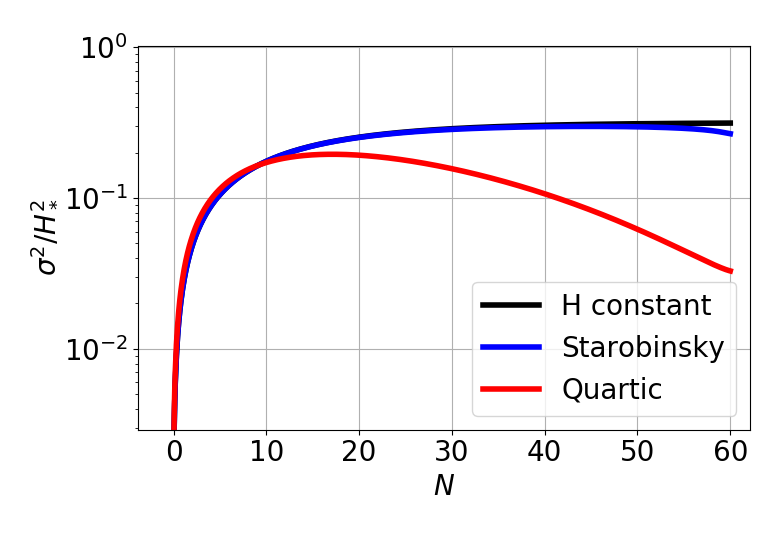}
	\caption{\textit{
	Impact of the time dependence of the Hubble scale $H$ on the evolution of the variance of the Higgs field.
	The three curves correspond to the solution \eqref{general-solution} with $f=1$, for different background evolutions: constant $H$, Starobinsky inflation and quartic inflation, all having the same value of the Hubble scale $H_\star$ 60 $e$-folds before the end of inflation, $\xi=-0.01$, and $C_5=C_6=0$ in \eqref{M2}.
	While in the plateau case the system almost relaxes towards the corresponding de Sitter equilibrium, 
	 the final variance differs by one order of magnitude in the quartic model.}}
	\label{varianceH}
\end{figure}
In comparison to previous works, we distinguish ourselves by evaluating the variance of the Higgs field on a time-dependent background. To emphasize its impact, here we consider the time dependence of $H$ alone, without including the effects of higher-order operators. The mass term $\mathcal{M}^2/H^2= -12\xi  (1-\epsilon/2)$ induced by the non-minimal coupling receives a small $\epsilon$ correction, but it becomes important only in the last $e$-folds of inflation. A much more important effect comes from the explicit time dependence of the noise term in \eqref{variance2}:  as $H$ decreases during inflation, the amplitude of the stochastic kicks also decreases, and the cumulative effects on the variance \eqref{general-solution} may be important, depending on the inflationary model. Obviously, one expects little deviation compared to the idealized description of 
constant $H$ in plateau models of inflation, in which $\epsilon=-\dot{H}/H^2$ is very small during the bulk of the inflationary evolution, to substantially grow only in the last $e$-folds. On the contrary, effects are more pronounced in models with a steady decrease of $H$, like in monomial inflation.

Eventually, note that taking the de Sitter equilibrium result \eqref{equilibrium}, with its parameters evaluated at time $N$, is not in general a good approximation to the full time-dependent result. As already noted before in a general context, when $H$ is evolving, this adiabatic equilibrium is a good approximation only if the relaxation time is smaller than the time scale over which $H$ varies \cite{Hardwick:2017fjo}. 
The latter is given in slow-roll inflation by $N_{H}=1/\epsilon$ so that the condition for approximate equilibrium becomes \begin{equation}
N_{\mathrm{rel}}\simeq \frac{H^2}{\M^2}<\frac{1}{\epsilon}=N_{H},
\end{equation} 
i.e. $\epsilon \lesssim \M^2/H^2$, where the right-hand side is $\lesssim {\cal O}(1)$ in situations with non-negligible stochastic effects. For single-field plateau models, this is not very constraining, given their very small values of $\epsilon$ in the bulk of the inflationary evolution. However, with $\epsilon=(H_{{\rm end}}/H)^{4/p}$ it is easy to show that the above condition is never satisfied in the relevant range $\M^2\lesssim H^2$ and for monomial inflation with an exponent $p>2$ \cite{Hardwick:2017fjo}. 
In particular, for this type of backgrounds, the PDF never reaches the de Sitter equilibrium associated to the time $N_*$, i.e. 
the one corresponding to the plateau reached in the approximation of constant $H$.

These expectations are confirmed by explicit numerical results in figure~\ref{varianceH}, where we show the exact solutions \eqref{general-solution} for the variance, in two examples which are representative of the above classes: Starobinsky and quartic inflation, normalized with the same initial value $H_\star$ for the Hubble scale. We used $f=1$ to focus on the effects of the time-dependence, we chose $\xi=-0.01$, and for comparison we display the corresponding solution with constant $H=H_\star$. The differences between Starobinsky inflation and $H={\rm const}$ are minor as they accumulate only in the last $e$-folds, whereas the time-dependence of the inflationary background has an important impact for quartic inflation, in which the final variance is comparatively decreased by one order of magnitude. We stress that such kind of effects is all the more important as the fate of the Higgs is exponentially sensitive to its variance, as we will see in section \ref{secmod1}.

The reader may wonder why we consider the model of quartic inflation, which is ruled out, for instance because it generates primordial gravitational waves with amplitude exceeding by far the observational constraints $r<0.07$ \cite{Akrami:2018odb}. The reason is that quartic inflation is ruled out in the sense of a single scalar field both driving inflation and generating primordial fluctuations. Here, on the contrary, we are only interested in the background dynamics, which governs the time-dependence of the Hubble rate, and hence the amplitude of stochastic effects. Curvaton-type or more general multifield models
may well have the same time-dependence of $H$ as quartic inflation, without being ruled out by constraints on $n_s$ and $r$, which depend on the precise mechanism at the origin of primordial fluctuations.

\subsubsection{Planck-suppressed derivative operators}
\label{planck_suppressed}
\begin{figure}[t]
	\centering
	\includegraphics[width=0.65\textwidth]{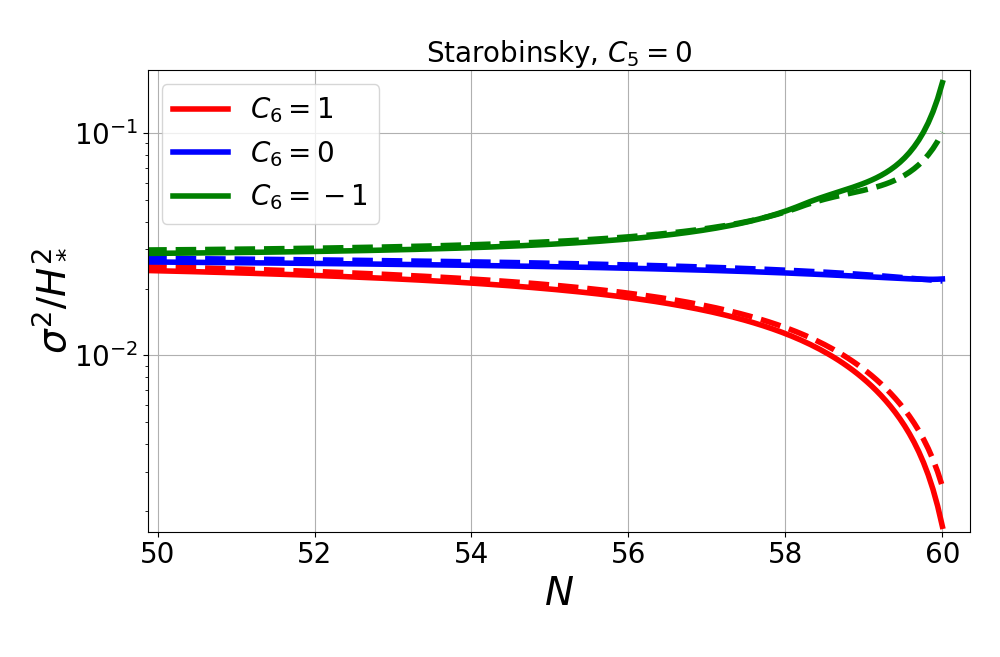}
	\caption{\textit{
	Impact of the dimension 6 derivative operator in \eqref{operatordes} on the evolution of the variance of the Higgs (normalized to $H_\star$). The three curves are for Starobinsky inflation, $\xi=-0.03$ and $C_6={1,0,-1}$, corresponding respectively to positive, vanishing and negative curvature of the field space manifold. For each case the full versus dashed lines represent the evolution determined by the conventional Fokker-Planck equation \eqref{Fokker} versus the phase-space one \eqref{W} discussed in \ref{fullstoca}. All curves are almost coincident for $N \leq 50$.}}
	\label{derivative-operators-plot}
\end{figure}
Let us now incorporate the effects of the derivative operators \eqref{kinetic-term}, which contribute to the effective mass of the Higgs as   
\begin{equation}\label{mass2}
\M^2/H^2 \simeq -12 \xi \left( 1-\frac{\epsilon}{2}\right)-3 C_5 \sqrt{2 \epsilon} \frac{\mpl}{M}+ 4 C_6 \epsilon \frac{\mpl^2}{M^2} \,.
\end{equation}   
As mentioned above, we concentrate on the minimal case of Planck-suppressed operators, i.e. $M=\mpl$. With $C_6$ and $C_5$ order one numbers, the dimension-five and 6 operators induce contributions to $\M^2/H^2$ of order $\sqrt{\epsilon}$ and $\epsilon$ respectively. These contributions are small $\Nmin$ $e$-folds before the end of inflation, although their contributions can already be similar to the one of the non-minimal coupling, depending on parameters and models. More importantly, their importance increases as inflation proceeds, with ${\cal O}(1)$ contributions by the end of inflation, at which $\epsilon=1$. Depending on the inflationary model and the signs of $C_6$ and $C_5$, the induced mass term can be positive or negative, and with a specific time-dependence, resulting in varied results. For simplicity, we only show in figure \ref{derivative-operators-plot} the evolution of the variance in a situation where effects are expected to be the least pronounced: with the dimension $6$ operator only (with $C_6=\pm 1$), and for Starobinsky inflation, for which $\epsilon$ substantially grows only in the last $e$-folds of inflation (see figure \ref{effect-b1} for the effect of the dimension-five operator). One can see that even in this situation, the effects of derivative operators are important, resulting in an increase (respectively decrease) of the final variance for a negatively (respectively positively) curved field space, with respect to the situation without these operators.
We note also that any contribution to the mass term, like the one of the derivative operators, has two combined effects, one deterministic and one stochastic, which go in the same direction: a positive contribution to $\M^2$ induces a steeper effective potential in the Langevin equation \eqref{langevin}, and a decrease of the amplitude of stochastic kicks, both further stabilizing the Higgs (a negative contribution acting in the other direction). We have checked that both of these (related) effects contribute substantially to the evolution of the variance.

\subsubsection{Stochastic formalism in phase space}
\label{fullstoca}

Since the effects of derivative operators emphasized in the previous section become increasingly important in the last $e$-folds of inflation, the reader might wonder if the assumption of a slow-roll evolution, or more precisely of an overdamped evolution, hidden in the Langevin equation \eqref{langevin}, is consistent. As we are going to show, taking into account the stochastic evolution in phase space does not modify significantly previous results.

In phase space, the evolution is described by two coupled Langevin equations, one for $h$ and one for its momentum $\pi \simeq \dot{h}=H dh/dN$, which can be written in general as
\begin{equation}
\label{system}
\frac{d X^{a}}{dN}=\h^a+g^a_\alpha \xi^\alpha, \quad X^a=\{h,\pi\},
\end{equation}
where $\xi^\alpha$ are independent normalized Gaussian white noises, verifying $\langle \xi^\alpha(N) \xi^\beta(N') \rangle=\delta^{\alpha \beta} \delta(N-N')$. In the situation of interest here, a test scalar field with quadratic potential, the amplitudes of the noises $g^a_\alpha$ do not depend on the $X^a$'s, i.e. the noises are not multiplicative. There is no It\^o versus Stratonovich ambiguity then \cite{vanKampen1981,Pinol:2018euk}, the $\h^a$ describe the deterministic dynamics \eqref{eom-h-approximate}, i.e. $\h^a=\{\pi/H,-(3\pi + \partial_h V_{\mathrm{eff}}/H)\}$, and the generalised FP equation for the probability distribution in phase space $W(h,\pi,N)$ reads 
\begin{equation}
\frac{\partial W}{\partial N}=\mathcal{L}(X) \cdot W, \quad \mathcal{L}(X)\equiv -\frac{\partial}{\partial X^a} \h^a +\frac{1}{2}\frac{\partial^2}{\partial X^a \partial X^b} D^{ab}\,,
\label{W}
\end{equation}
where the diffusion coefficients $D^{ab}=\delta^{\alpha \beta}  g^a_\alpha g^b_\beta$ are nothing else than the correlation functions of the UV modes of $h$ and $\pi$ when they reach the IR sector at $k=w aH$. Similarly as above, we take as initial conditions a Dirac distribution in phase space $W(h,\pi,0)=\delta(h) \delta(\pi)$. As the dynamics is still linear, $W$ subsequently follows a centered Gaussian distribution in phase space, whose evolution of the variances is simply obtained from Eq. \eqref{W} as
\begin{equation}\label{systemfull}
\begin{split}
\frac{\partial\langle h^2\rangle}{\partial N}&=\frac{2}{H}\langle h \pi\rangle+ D^{hh},\\[1mm]
\frac{\partial\langle h \pi\rangle}{\partial N}&=-\frac{\M^2}{H}\langle h^2\rangle -3\langle h\pi\rangle +\frac{1}{H}\langle \pi^2\rangle+D^{h \pi}\\[1mm]
\frac{\partial\langle \pi^2\rangle}{\partial N}&=-\frac{2\M^2}{H}\langle h \pi\rangle -6\langle \pi^2\rangle +D^{\pi \pi}\,.
\end{split}
\end{equation}
Different prescriptions for the diffusion coefficients have been discussed in the literature \cite{Riotto:2011sf,Rigopoulos:2016oko,Moss:2016uix,Prokopec:2017vxx,Grain:2017dqa,Pinol:2018euk,Cruces:2018cvq,Pattison:2019hef}. In figure \ref{derivative-operators-plot}, we used the simple one $D^{hh}=(H f/2 \pi )^2$, with the other coefficients vanishing, corresponding to neglecting the stochastic kicks of the momentum $\pi$. It is apparent that the effects of derivative operators and of the time-dependence of $H$ we are interested in are well described by the conventional stochastic approach \eqref{Fokker} of previous sections. In other words, considering the difference between the conventional field-space approach and the phase-space one as a measure of the theoretical uncertainty of phase-space effects, we can see that the latter is negligible for our purpose.\footnote{While we do not fully understand the motivations from the authors for this prescription, we have also checked that we obtain very similar results when using $D^{\pi \pi}=(3 H^2 f/2 \pi)^2$, with other coefficients put to zero, which is advocated in \cite{Rigopoulos:2016oko,Moss:2016uix,Prokopec:2017vxx} (with $f=1$). This can be seen as a further proof that the phase-space theoretical uncertainty is negligible for our purpose.} Eventually, while figure \ref{derivative-operators-plot} considers the effect of the dimension-six operator, the same conclusion is reached for the dimension-five operator, see figure \ref{effect-b1}. As a result, in the rest of the paper, we stick to the conservative stochastic approach described by the Fokker-Planck equation \eqref{Fokker}.


\section{Probability of falling in the AdS vacuum} \label{section3}

In this section we explain our procedure to extract, from the evolution of the PDF $P(h,N)$ studied in section \ref{stochastic}, the fraction of AdS patches that can reside in our past light cone. A few precautionary words: the word fraction is used in a probabilistic sense here, and the approach used in this paper, as any study of the Higgs stability during inflation, relies on sampling something that is by definition unique, i.e. our observable universe. As a consequence, if not satisfied, the inequality in Eq. \eqref{prob} below would not necessarily imply that our universe can not exist with these initial conditions and parameters, but would tell us instead that it is very unlikely.

\subsection{No AdS patch in our past light cone}\label{anthropic}

We label with $\mathcal{F}_{\mathrm {AdS}}$ the fraction of patches in AdS at the end of inflation.
Following
Ref.~\cite{Espinosa:2015qea}, the existence of our universe as we know it, with the Higgs in the electroweak vacuum, requires the following bound to be satisfied:
\begin{equation}\label{prob}
\mathcal{F}_{\mathrm {AdS}}\times \mathcal{N}<1,
\end{equation}
where $\mathcal{N}$ represent the number of Hubble patches present at the end of inflation in the volume giving rise to our observable universe today,
 i.e. $\mathcal{N}=H_0^{-3}/(a_0H^{-1}_{\mathrm{end}}/a_{\mathrm{end}} )^3\simeq e^{3N_\star}$. 
We might as well be interested in the fraction of patches that can potentially lead to AdS regions, despite being still safe at the end of inflation.
The fate of Hubble patches with values of the Higgs greater than the location $h_{\mathrm{max}}$ of the potential barrier depends on the details of the post-inflationary dynamics (see Ref.~\cite{Espinosa:2015qea} for details and Ref. \cite{Jain:2019wxo} for a scenario in which reheating does not happen instantaneously).

After inflation, the Higgs potential receives thermal corrections from the SM bath, contributing to the Higgs mass as \cite{Giudice:2000ex}:
\begin{equation}
\mathcal{M}_T^2\simeq T^2(a)e^{-h^2/(2\pi T)^2},
\label{M4}
\end{equation}
with $T(a)=1.3 \,T_{\rm {m}}a^{-3/8}(1-a^{-5/2})^{1/4}$ and $T_{\rm{m}}=0.54\, (0.1 H_{\rm {end}}\mpl \Trh)^{1/4}$, where $\Trh$ is the reheating temperature, and we set $a=1$ at the end of inflation. The thermal contribution adds to the one coming from the mass $\mathcal{M}^2$ induced by the non-mimal coupling and the higher-order operators. This term after inflation becomes
 \begin{equation}
 \mathcal{M}^2\simeq \left(-\xi -\sqrt{3}C_5+2 C_6\right) \frac{3 H^2_{\rm {end}}}{a^3} \,,
 \label{M3}
 \end{equation}
where we have assumed that, while the inflaton is oscillating about the minimum of its potential, the Universe experiences a matter dominated phase so that $H=H_{\rm {end}}/a^{3/2}$ and $\epsilon = 3/2$. These equalities have to be thought as the result of averaging over many oscillations. 
Thus, the rescuing ability of these corrections depends on the reheating temperature $T_{\mathrm{RH}}$ as well as on $C_5,C_6$ and $\xi$. The interplay between the decay of the mass contributions \eqref{M4} and \eqref{M3} and the dynamics of the Higgs during this stage determines if a patch with a given value $h$ is brought back to the safe region $|h|<h_{\rm{max}}$ before $\mathcal{M}+\mathcal{M}_T$ becomes negligible.
In short: any given set of parameter corresponds to a maximum value of $h$, usually labelled as $h_{\rm {end}}$, that can be rescued and brought back safely to the EW vacuum. As can be seen from \eqref{M3}, our operators during the post-inflationary phase are relevant, shifting $\xi$ by an order one number, i.e. $\xi\rightarrow \xi_{\rm{eff}}=\xi+\sqrt{3}C_5  -2C_6$. One could thus follow the same procedure as in \cite{Espinosa:2015qea}, but now with a new effective $\xi$, to determine $h_{\rm {end}}$.

For sufficiently large reheating temperature the thermal corrections always dominate and for
approximately $ \Trh \gtrsim 10^{13}\,\mathrm{Gev}$, patches (which are not yet in AdS) with arbitrary large values of the Higgs can be rescued. Conversely, for low reheating temperature $\Trh\lesssim 10^{5}\mathrm{GeV}$, and no induced mass coming from the additional operators, i.e. $\xi=C_5=C_6=0$, any patch in which $|h|>\hmax$ will end up forming an AdS region. Thus, less conservative bounds can be derived by asking that there was no patch of that type at the end of inflation: 
\begin{equation}\label{prob2}
\mathcal{F}_{ \mathrm{|h|>\hmax} } \times \mathcal{N}<1,
\end{equation}
with $\mathcal{F}_{\mathrm {|h|>\hmax}}$ the fraction of Hubble regions where $|h|>\hmax$. 

It is worth mentioning that even for negligibly small thermal corrections, but $(\xi,C_5,C_6)\neq0$,
the maximum value of the Higgs that can be rescued thanks to the post-inflationary dynamics ($h_{\rm {end}}$) is indeed different from $h_{\rm {max}}$.\footnote{In particular, in Ref. \cite{Espinosa:2015qea}, it is shown, for the illustrative case  $\xi\sim -0.1$ (value in the range of interest for stochastic kicks to be effective during inflation), that the rescuing ability of \eqref{M3} becomes relevant for $T_{\rm {m}}\sim h_{\rm {max}}$ and $H_{\rm {end}}/h_{\rm {max}}\gtrsim 10^2$ (see their Fig. 9). In our case, we expect the rescuing effect to be amplified and to become relevant even for higher reheating temperature, or analogously, to provide the same effect in absence of thermal corrections as the one given by higher reheating temperature. This expectation is motivated by the shift of order one that the higher-order operators induced on $\xi$ in \eqref{M3}.}
On top of that, the exact determination of $h_{\rm{end}}$ also depends on how the reheating phase is modelled. Thus, given that the main interest of this work concerns the dynamics during inflation, we show results only for the two bounds \eqref{prob} and \eqref{prob2}, corresponding to cases where the impact of the post-inflationary dynamics is maximal and minimal respectively.

\subsection{Matching stochastic and classical dynamics}
\label{matching}
The picture of a spectator Higgs field undergoing a stochastic motion and subject to a quadratic potential is (obviously) not the right description in a patch that is falling towards the true vacuum and forming an AdS region.
At large enough field values, the effect of the quadratic SM potential is not negligible anymore, so that the PDF becomes non-Gaussian, and more importantly, the backreaction of the Higgs on spacetime becomes important.
Given the complexity of the system, all approaches used to model it should rely on some approximation schemes.
Thus, before illustrating our procedure, we find it convenient to discuss different ones present in the literature. 

First, let us define $\hcl$ as the Higgs value at which the dynamics becomes classically dominated. 
It can be estimated by requiring that the deterministic part driven by the effective potential in the Langevin equation \eqref{langevin} overcomes the 
noise term:\footnote{In Ref.~\cite{Espinosa:2015qea}, $\hcl$ 
is determined in an almost equivalent way by considering where the deterministic term overcomes the stochastic one in the FP equation. This gives a slightly different result which does not affect our conclusions. We prefer to use Eq. \eqref{had} to determine $\hcl$ as it is independent of the Gaussian ansatz for the PDF, ansatz that precisely breaks down around $\hcl$.}
\begin{equation}\label{had}
h_{\mathrm cl}:\quad\Bigg| \frac{\partial_h V_{\mathrm{eff}}}{3 H^2} \Bigg|=\frac{H f}{2\pi},
\end{equation}
and $\hcl$ is such that if $h\gtrsim \hcl>\hmax$ the Higgs will classically roll towards the true vacuum.
As a first approximation, it was assumed in Ref.~\cite{Espinosa:2015qea} that once the Higgs reaches $h_{\rm{cl}}$, it instantaneously forms an AdS region. This way of proceeding brings some important simplifications. Since up to $|h|\lesssim \hcl$ the contribution coming from the SM quartic potential can be neglected to a good approximation, one can model the PDF with a Gaussian in the bulk region $[-\hcl,\hcl]$, and 
cut its tails at $|h|\geq\hcl$. Then, the fraction of patches in AdS can be estimated by computing $\mathcal{F}_{\rm AdS}=1-{\cal P}(|h|<\hcl)$ at the end of inflation.
Later, in Ref.~\cite{East:2016anr}, the finite time to fall in AdS from $\hcl$ has been taken into account in the following way: the FP equation \eqref{Fokker} was used beyond the value $\hcl$, although the noise becomes negligible then, which enables one to capture the non-Gaussian tails induced by the classical effects of the SM potential. This procedure was applied up to the value $\hnsr$ where the backreaction of the Higgs was estimated to generate very rapidly an AdS region. Similarly as above, the fraction of AdS patches was then computed as $\mathcal{F}_{\rm AdS}=1-{\cal P}(|h|<\hnsr)$ at the end of inflation, returning results similar to the Gaussian approximation of Ref.~\cite{Espinosa:2015qea}, which is expected in situations when a steady PDF is reached much before the end of inflation.

We are now in the position to highlight one of the key differences compared to previous works. In all previously studied setups (at least to the best of our knowledge), the effective potential is a static function during inflation. As a consequence, the fraction of patches transmuting in AdS can only increase during inflation, i.e. the PDF initially peaked at
zero can only flatten while its tails become fatter. 
As we discussed in section \ref{secevo}, the effects we are interested in, from higher-order derivative operators and de Sitter departure, are genuinely time-dependent, and they can act as rescuing processes shrinking the PDF while inflation proceeds. Therefore, we have to carefully take into account that patches  once in AdS cannot be brought back to the safe region for the Higgs field. 
In other words, we have to model a distribution that is losing part of its tails, and which later cannot be re-introduced in the bulk when the distribution following the FP equation is shrinking.\footnote{In a static situation, an early work on the subject analytically solves the FP equation with boundary conditions that act as sinks at some given Higgs values \cite{Espinosa:2007qp}. This is not possible in our case, because of the time dependence of the background, together with the fact that the point where the Higgs is trapped in AdS changes with time. Attacking this problem fully numerically is an interesting option, but goes beyond the scope of this work.}
Furthermore, the time dependence introduced by our analysis modifies on a case by case basis the time to fall in the AdS vacuum, as well as the value of $h_{\rm cl}$ computed from \eqref{had}. Thus, for any given setup (model for the evolution of $H$, value of the non-minimal coupling and values of the Wilson coefficients for the derivative operators) we proceed in the following way:
\begin{itemize}
	\item We model the evolution of the PDF with a Gaussian satisfying the FP equation \eqref{Fokker} up to the value $|h|=\hcl$. The point of classicality computed from Eq.~\eqref{had} changes over time in a different way for each framework. For instance, the decrease
	of $H$ dynamically extends the range of the classical region.
	Implicitly we can write
\begin{equation}
	h_{\mathrm cl}(N)\equiv h_{\mathrm cl}(H(N),\xi,C_6,C_5,V_{{\rm SM}}).
	\end{equation}

	\item In the classical region, we numerically trace the evolution of the full two-field inflaton-Higgs system, including the backreaction of the Higgs
	on the background. We consider the fall
	in AdS unavoidable when the Hubble parameter (in the Einstein frame) becomes negative, i.e. $H_E<0$.\footnote{In our simulations this happens less than one $e$-fold after the time where the full potential crosses zero, i.e. $\Vinf(\phi)+V_{\mathrm {SM}}\simeq 0$.}
	At each time $N$ of the evolution we compute the number of $e$-folds $N_{\mathrm {AdS}}$ necessary to fall in AdS with initial conditions given by $h=\hcl(N)$. We say that
\begin{equation}\label{settime}
	\hcl(N)\in \mathrm{AdS}\,\,\,\,\,\mathrm{if}\,\,\,\,\,N_{\mathrm {AdS}}<\Nmin-N\,,
	\end{equation} 
	where $\Nmin-N$ is the number of $e$-folds left before the end of inflation when the Higgs starts its classical dynamics at the value $\hcl(N)$.
	
	\item We estimate the fraction of patches in AdS at the end of inflation by computing the maximum area of the distribution under the tails, namely for $|h|>\hcl(N)$, 
amongst the times $N$ such that $\hcl(N)\in \mathrm{AdS}$, i.e. that leave enough time before the end of inflation for a patch of value $\hcl(N)$ to fall in AdS.
		In order to exclude the possibility of AdS patches in our past light cone, we impose the bound from Eq. \eqref{prob}: 
\begin{equation}\label{prob-result}
	\mathcal{F_{\mathrm {AdS}}}=    \max_{\{N\,:\,\hcl (N)\in \mathrm{AdS}\}}\left[\mathcal{P}(|h|>h_{\rm cl}(N),N)\right]=\mathcal{P}(|h|>h_{\rm cl}(\Nm),\Nm)<e^{-3\Nmin}\,,
	\end{equation}
where we call $\Nm$ the time at which the maximum is evaluated. Because of the time dependence of $\hcl(N)$ and the finite time to fall in AdS from there, it is worth stressing that the maximum is not necessarily reached at the time when the variance of the PDF has grown to its largest value.

	\item For the purpose of estimating the fraction of patches where $|h|>\hmax$ at the end of inflation, we 
	take the PDF evolved until then, i.e. $P(h,N_{\mathrm {end}})$. This can be used on the condition that 
	\begin{equation}
	\mathcal{P}(|h|<h_{\mathrm {max}},N_{\mathrm {end}})<\mathcal{P}(|h|<h_{\mathrm {cl}},N_{\mathrm {m}})\equiv1-\mathcal{F}_{\mathrm {AdS}}.
	\end{equation}
This is imposed to exclude (approximately) patches that are judged safe at the end of inflation according to the FP evolution alone, but that have actually classically fallen in AdS before. As an approximate way of taking into account these AdS tails that can not be rescued, and following Eq. \eqref{prob2}, we therefore impose the bound
	\begin{equation}
	\mathcal{F}_{|h|>h_{\rm max}} =  {\max} \left[ \mathcal{P}(|h|>h_{\mathrm{ max}},N_{\mathrm {end}}) , \mathcal{F_{\mathrm {AdS}}}  \right]< e^{-3\Nmin}
	\label{prob2-result}
	\end{equation}
	in order to exclude the possibility of patches with $|h|>h_{\mathrm {max}}$ in our past light cone.

\end{itemize}
Before moving on, it is worth mentioning another possibility one can in principle follow to perform the analysis. The reader may indeed wonder why one does not simply sample a large number of evolutions using the Langevin equation with initial conditions given by $h=0$.
Then fit the distribution of the final Higgs values at the end of inflation with a PDF and with that compute the survival probability. Unfortunately, proceeding in this way would overestimate the impact of the effects studied in this paper. Indeed, given the smallness of the probabilities we are considering (see Eq.~\eqref{prob}), any reasonable number of realizations would always 
probe the central part of the distribution. This can be correctly and safely extrapolated to compute the tails in a static case, like in Ref.~\cite{Franciolini:2018ebs}. However, in our situations, it would return the PDF evolved with the FP up to the end of inflation. As already mentioned, this distribution 
ignores the important fact that patches which are in AdS at a given time during inflation cannot later on be brought back to the safe region for the Higgs, and therefore is not trustworthy.

\subsection{Analytical considerations}\label{secmod1}

Before moving to the full numerical results it is useful to draw a few analytical considerations about the bounds \eqref{prob-result}-\eqref{prob2-result}.
As discussed in the previous section, the PDF can be approximated with a Gaussian distribution. Thus, from Eq.~\eqref{prob-result}, we obtain 
\begin{equation}\label{appro}
\mathcal{F}_{\mathrm {AdS}}\equiv\mathcal{P}(|h|>\hcl(\Nm),\Nm)=1-\text{erf}(x)\simeq \frac{e^{-x^2}}{\sqrt{\pi}x}<e^{-3\Nmin},\quad\quad x\equiv\frac{\hcl(\Nm)}{\sqrt{2}\sigma(\Nm)}.
\end{equation}
Given the very small probability $e^{-3\Nmin}$ we are considering, with $\Nmin \simeq 60$, the approximation of the error function is robust, and $\Nm$ can be estimated by minimizing $\hcl(N)/\sigma(N)$ within the domain where the Higgs has enough time to fall in AdS before inflation ends, see Eq. \eqref{settime}. If the time dependence of $H$ was ignored, $\hcl$ would be constant, and
$\Nm$ would occur at the maximum of $\sigma$. This is not our case though, and different possibilities can arise. For instance, 
in situations where $\hcl$ decreases and $\sigma$ grows, $\Nm$ is  
simply the latest time at which $\hcl(N)\in \mathrm{AdS}$. If both $\hcl$ and $\sigma$ decrease, the competing effect between the evolution of the classicality point and the one of the variance determines whether a given setup alleviates or worsens the Higgs instability.\\
To estimate $\hcl$ from
Eq. \eqref{had} we first approximate the SM potential as \cite{Espinosa:2015qea}
\begin{equation}\label{appro2}
V_{\mathrm{SM}}\simeq -b \ln\left(\frac{h^2}{h^2_{\mathrm{max}}\sqrt{e}}\right)\frac{h^4}{4},
\end{equation}
with $b\simeq 0.16/(4\pi)^2$ for central values of the SM parameters. 
When only the SM potential is present, Eq. \eqref{had} can be solved exactly:
\begin{equation}\label{h4}
h^{(4)}\equiv\(\frac{\alpha f H^3}{\mathcal{W}\[\frac{\alpha H^3}{h^3_{\mathrm{max}}}\]}\)^{1/3},
\end{equation} 
with $\alpha\simeq9\cdot4\pi/0.16$ and where $\mathcal{W}$ is the Lambert function  (or product logarithm)  function defined as the inverse function of $f(y)=y e^y$, i.e. $z=\mathcal{W}(z)e^{\mathcal{W}(z)}$.
In the presence of a quadratic term in the effective potential, that we write
schematically as 
$V_{\mathrm {eff}}=V_{\mathrm{SM}}+V^{(2)}$, 
Eq. \eqref{had} has no exact solution.
However, $h_{\mathrm {cl}}$ is well approximated by the value $\tilde{h}$ at which
\begin{equation}\label{tildeh}
|\partial_h V_{\mathrm {SM}}|=\partial_h V^{(2)} \implies \tilde{h}=\left(\frac{\mathcal{M}^2}{b \mathcal{W}\[\frac{\mathcal{M}^2}{b h^2_{\mathrm {max}}}\]}\right)^{1/2},
\end{equation}   
if $\tilde{h}>h^{(2)}$, where $h^{(2)}$ is such that $\partial_h V^{(2)}=3H^3f/2\pi$, which means that the quadratic term dominates at small field values. 
Analogously if $\tilde{h}<h^{(2)}$, which happens for very small masses so that $V^{(2)}$ 
is negligible, we can use \eqref{h4} to approximate $h_{\mathrm {cl}}$. 
Considering for definiteness the case where $h_{\mathrm {cl}}\simeq \tilde{h}$ in \eqref{appro}, we arrive at the bound
\begin{equation}
{\rm No\,AdS\,patch:} \qquad
\frac{H(\Nm)}{h_{\mathrm{max}}}<\frac{1}{\sqrt{6 \Nmin \hat{\sigma}^2}}\, e^{\beta/ \hat{\sigma}^2},\quad\quad \hat{\sigma}^2\equiv \frac{\sigma^2(\Nm)}{H^2(\Nm)}\,,
\label{bbound}
\end{equation}
with $\beta=\mathcal{M}^2/ (12H^2 \Nmin b)|_{N=\Nm} $. This generalizes the one used in Ref.~\cite{Espinosa:2015qea}, to which it reduces when assuming a constant Hubble scale and the variance given by the equilibrium solution in Eq. \eqref{equilibrium}. It is worth stressing the exponential factor in the right-hand side, which renders the bound above very sensitive to even small changes in $\hat{\sigma}^2$ and $\mathcal{M}^2/H^2$. This is indeed very important as we have seen that the various effects studied in this paper have in general a substantial impact on these quantities, and notably on the (normalized) variance of the Higgs.

Finally we discuss the bound \eqref{prob2-result}
coming from the request of avoiding patches with $|h|>h_{\mathrm {max}}$. With precautions spelled out in Sec. \ref{matching}, one may use for this purpose the variance evaluated at the end of inflation, which, analogously to Eq. \eqref{appro}, gives rise to
\begin{equation}
{\rm No\,patch\,over\,the\,barrier:} \qquad
\frac{H_{\star}}{h_{\mathrm{max}}}<\frac{1}{\sqrt{6\Nmin \hat{\sigma}^2_{\mathrm{end}}}},\quad\quad \hat{\sigma}^2_{\mathrm{end}}\equiv\frac{\sigma^2(N_{\mathrm{end}})}{H^2_\star} \,,
\label{bbbound}
\end{equation}
where we normalized the variance with respect to $H_{\star}$, the value of the Hubble scale when the pivot scale exits the Hubble radius. This bound is less sensitive to changes of the variable $\hat{\sigma}^2_{\mathrm{end}}$. Nevertheless, we have seen that the latter may vary by orders of magnitude when varying the inflationary model (at fixed $H_\star$), or details of Planck-suppressed operators, which renders this bound also an interesting probe of these aspects.


\section{ 
Numerical Results}
\label{secres}

When approximating inflation with an exact de Sitter phase,
and neglecting possible contributions from Planck-suppressed operators, the fate of the Higgs only depends on the SM parameters measured at the EW scale, the energy scale during inflation $H_\star$, and the value of the non-minimal coupling $\xi$.\footnote{As already discussed, we are not interested in the post-inflationary dynamics, which would introduce at least one extra parameter dependence through the reheating temperature.
In this respect, the two bounds \eqref{prob} and \eqref{prob2} should be thought of as the two 
extreme limits of the effects of thermal corrections to the Higgs potential 
after inflation, one where there is no rescuing effect ($T_{\mathrm{RH}}\lesssim 10^5\,\mathrm{GeV})$, and the other with maximum rescuing effects ($T_{\mathrm{RH}}\gtrsim 10^{13}\,\mathrm{Gev}$).} Motivated by our study in Sec.~\ref{stochastic}, our aim in this section is to exemplify the additional sensitivity of the fate of the Higgs on the time dependence of the inflationary background, and on derivative Planck-suppressed operators. 

For the latter, we vary the two Wilson coefficients $C_5$ and $C_6$ in Eq.~\eqref{kinetic-term} in the range $\{-1,0,1\}$. To model the time evolution, we use two different background dynamics. The first is given by a plateau type inflationary potential \`{a} la  Starobinsky \cite{Starobinsky:1980te}, which gives rise to an evolution for the Hubble rate as $H(N)\simeq H_*\exp[O(1)/(N-N_{\mathrm{end}})]$. In that case, $\epsilon$ and $H$ are nearly constant when the inflaton evolves along the plateau of the potential, and $H$ changes only by an order one factor in the last $e$-folds.
In contrast, in monomial potentials these quantities have a non-negligible evolution throughout the inflationary phase (as we have seen, with the consequence of the system never reaching the de Sitter equilibrium for monomial inflation with exponent greater than $2$),
e.g. the Hubble rate evolves as $H(N)=H_* (1+N_{\mathrm{end}}-N)/(1+N_{\mathrm{end}})$ for a dynamics \`{a} la quartic. In our numerical results, we thus make use of the inflationary potentials $V(\phi)=\Lambda^4(1 - e^{-\sqrt{2/3}\phi/\mpl})^2$ and $V(\phi)=\lambda \phi^4$. 
In all scenarios, we take into account the deviation of the amplitude of the stochastic noise from the massless limit, as set by the function $f$ in Eq. \eqref{f-explicit}. Different results obtained for the two background evolutions highlight one of our main points: the bounds are sensitive to the de Sitter departure and are therefore inherently model-dependent.

Following the procedure explained in Sec.~\ref{matching}, we numerically computed the fraction of AdS patches at the end of inflation, and the fraction of patches in which the Higgs has fluctuated above the potential barrier, for different scenarios. We present these results by showing constraints on the energy scale of inflation in section \ref{secsur}, and on the SM parameters in section \ref{secbou}, both coming from the requirement of not having a single patch in AdS in our past light cone.

\subsection{Bounds on the energy scale of inflation}
\label{secsur}

\begin{figure*}[t!]
		\hspace{-0.55cm} 
	\begin{subfigure}[t]{0.5\textwidth}
		\centering

		\includegraphics[width=1.0\linewidth]{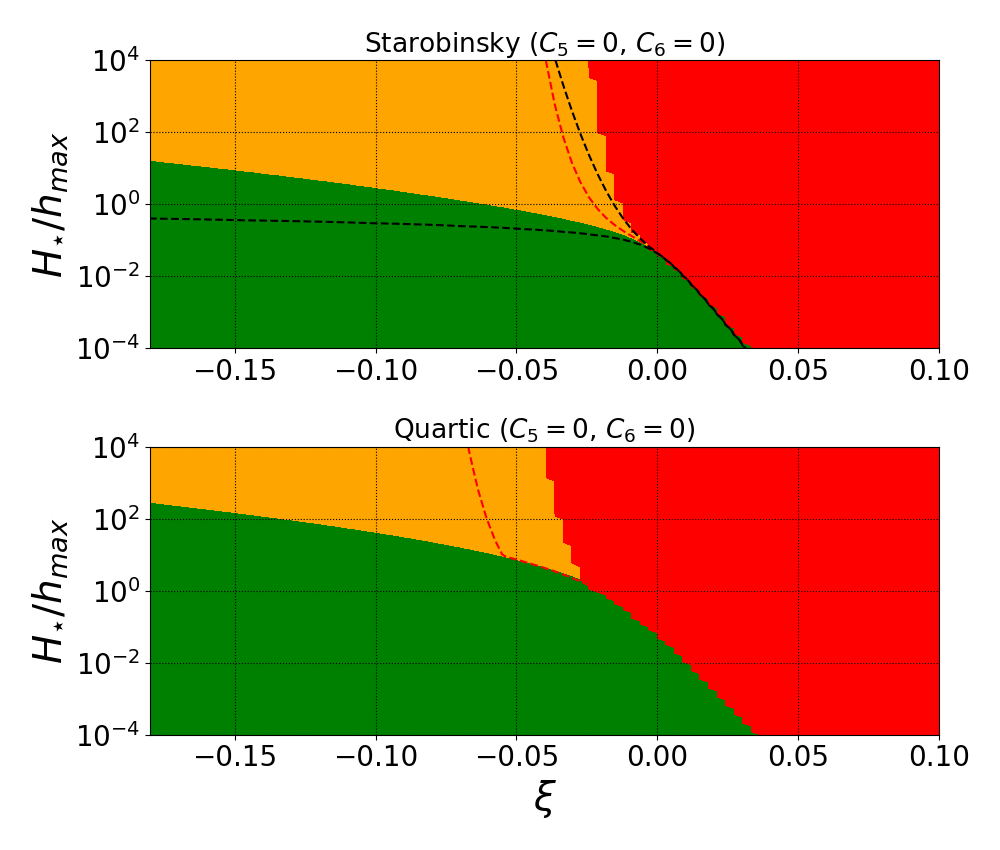}
		\caption{\textit{Fate of the Higgs without derivative operators, for Starobinsky-like and quartic-like inflationary evolutions.}}
		\label{survival1}
	\end{subfigure}%
	~ \hspace{0.55cm}
	\begin{subfigure}[t]{0.5\textwidth}
		\centering
		\includegraphics[width=1.0\linewidth]{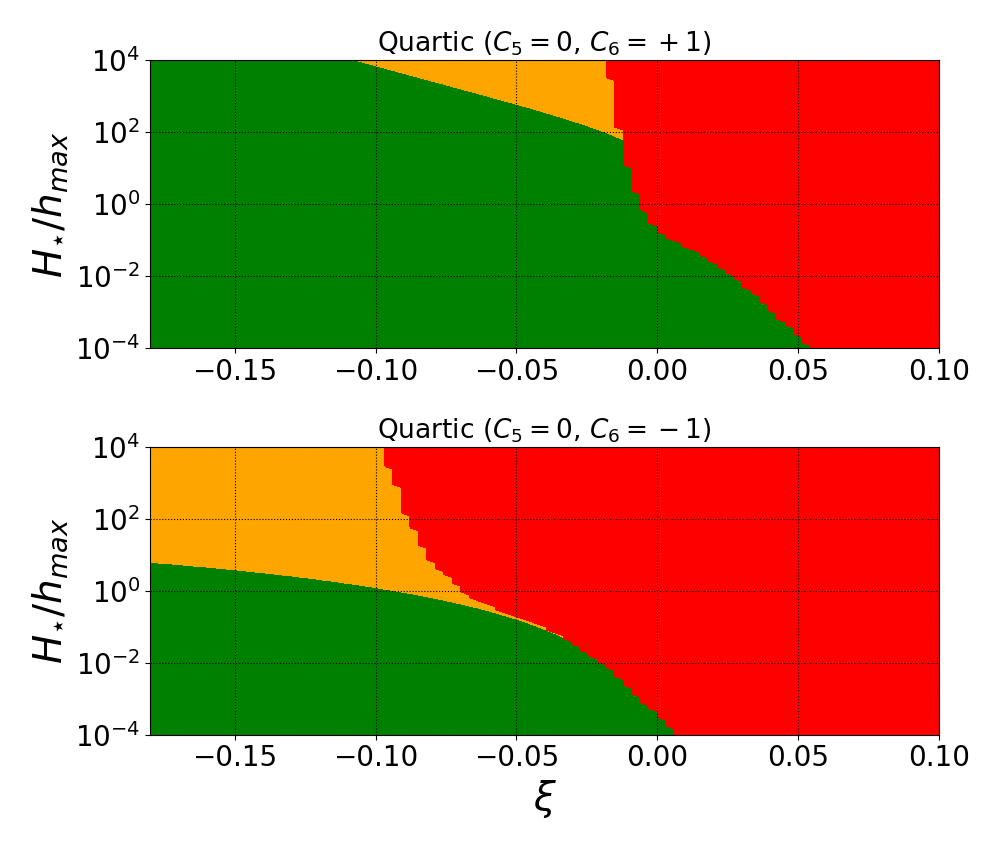}
		\caption{\textit{Impact of the dimension-six operator on the fate of the Higgs for quartic inflation.}}
		\label{survival2}
	\end{subfigure}
	\caption{\textit{Cosmological fate of the Higgs for different evolutions of the scale factor and parameters. 
		The green region represent scenarios where there is not a single Hubble patch in our past light cone in which the Higgs has fluctuated above the barrier. The orange region represents the potentially unsafe scenarios, in which there is no patch in AdS at the end of inflation, but there exist patches above the barrier that can potentially become AdS regions depending on the post-inflationary dynamics. In red: the region in which there exists at least one Hubble patch in AdS at the end of inflation, and the corresponding model cannot describe our observable universe. The red dashed curved on the left plots highlights the would-be boundary of the red region if the finite time to fall into AdS was not taken into account. All plots are obtained considering the noise in the stochastic process given by the function $f$ in Eq.~\eqref{f-explicit}. The dashed black lines on the top left plot mark off the boundaries between the three regions for the benchmark analysis  
		assuming $H$ constant and $f=1$.}}
	\label{survival}
\end{figure*}

For each scenario that we consider, i.e. a given evolution of the scale factor during inflation and precise Planck-suppressed couplings, we compute
$\mathcal{F}_{\mathrm{AdS}}$ and $\mathcal{F}_{\mathrm{|h|>h_{max}}}$ for different values of the non-minimal coupling $\xi$ and of $H_\star/h_{\mathrm{max}}$.
The results are displayed in figure \ref{survival}, where the three coloured regions, following Ref.~\cite{Espinosa:2015qea}, are defined as follows:

\begin{itemize}
	\item The red region is the part of parameter space where there is (on average) at least one Hubble patch in AdS at the end of inflation, hence the corresponding model cannot describe our observable universe. 
	\item In the orange region, at least one patch has fluctuated above the potential barrier at the end of inflation, but without falling into AdS. These patches may or may not be rescued depending on the post-inflationary dynamics, and in particular the reheating temperature. Thus, we label this region as \textit{potentially unsafe}. 
	\item In the green region, not a single patch has fluctuated above the barrier during inflation, i.e. both bounds \eqref{prob} and \eqref{prob2} 
are satisfied, and the Higgs safely rolls towards our electroweak vacuum after the end of inflation.
\end{itemize}

\subsubsection{de Sitter departure}

In figure \ref{survival1}, we first look at the time-dependent effects alone, so that we set to zero both $C_5$ and $C_6$ in the Higgs effective potential. In black dashed lines we highlight the boundaries of the green-orange-red regions under the assumptions of a de Sitter background ($H=$ constant) and noise amplitude
given by the one of an exactly massless scalar field, i.e setting $f=1$. Our motivation is to provide a direct comparison between our results and 
previous ones present in the literature \cite{Espinosa:2015qea}.  
In this respect, one can
notice the significant shift of the boundary at large negative values of $\xi$. This comes from the increased mass of the Higgs fluctuations, which results in the suppression of the amplitude of the noise, i.e. $f <1$.
Hence, the transition to the region in parameter space where the stochastic noise is irrelevant now appears smoothed.

The dashed red line indicates the boundary of the would-be red region if we had ignored the finite time to fall into AdS, 
i.e. if we had simply considered the probability to fall to AdS at its maximum value during the evolution.
The difference with the actual boundary of the red region highlights the importance, in our approach, to consider the classical evolution of the two-field system. Since the time to fall into AdS is a non-trivial quantity in a time-dependent background, and given that it is 
highly setup-dependent, we provide further details about it in appendix \ref{time-ads}.

In order to better understand the differences between the two models in figure \ref{survival1}, let us recall that the time dependence of $H$ leads to two effects that play a role in determining the fate of the vacuum instability, and that may compete: the decrease of the variance (see Sec.~\ref{Htimedependence}), and the time dependence of $h_{\mathrm{cl}}$, the point after which the dynamics becomes classically dominated (see Eq.~\eqref{had}). When the variance decreases, the probability of being beyond an arbitrary fixed value of the Higgs diminishes. Yet, $h_{\mathrm{cl}}$ is a dynamical quantity that may decrease at such a rate that compensates for this effect, resulting in a net increase of $\mathcal{P}(|h|>h_{\mathrm{cl}})$.

For Starobinsky-like evolution, as discussed in Sec.~\ref{secevo}, for sizeable enough values of $\xi$, the variance initially reaches the value of the de Sitter equilibrium associated to $H_\star$. Afterwards, the variance and $h_{\mathrm{cl}}$ decrease with the net effect that the probability of being beyond $h_{\mathrm{cl}}$ increases.
However, this occurs only in the last $e$-folds of inflation, so that patches with values around $h_{\mathrm {cl}}$ do not have the time to fall in AdS. Hence, in this situation, our careful way \eqref{prob-result} of computing $\mathcal{F}_{\mathrm{AdS}}$, 
which determines the position of the red-orange boundary, does not lead to an important difference compared to de Sitter zeroth-order result (dashed black line). The only notable difference 
concerns the boundary between the allowed (green) and potentially unsafe
 (orange) region: it is lifted due to the suppression of the noise coming from the function $f$, and the decrease, as $H$ diminishes, of the probability of being beyond the fixed value $h_{\mathrm{max}}$.

The larger rate of change of $H$ in quartic-like evolution, despite reducing the variance at a greater rate, results in an expansion of the disallowed (red) part of parameter space compared to plateau-like models. Conversely, as $h_{\mathrm{max}}$ is a fixed point, the quartic evolution acts as rescuing when we look at the boundaries between the green-orange regions.
There, the shrinking of the variance leads to recovering patches that would have been otherwise beyond the potential barrier at the end of inflation.

More intuitively, one can also understand the above physical consequences of the time-dependence of $H$ as follows.
The decrease of $H$ determines a smaller size of the random kicks of the Higgs field. Thus, regions where the Higgs is just above the potential barrier can be more easily rescued under the same positive mass term (as the one induced by the non-minimal coupling $\xi$). This leads to a larger green region for quartic inflation in figure \ref{survival1}. At the same time, regions where the Higgs has fluctuated far away beyond the potential barrier could not be rescued anymore given the smallness of the quantum jumps. These regions become effectively classically dominated, with the only option to fall into AdS. This leads to a larger red region for quartic inflation in figure \ref{survival1}.

\subsubsection{Planck-suppressed derivative operators}

In figure \ref{survival2}, we show the effects of the dimension-six Planck-suppressed operator for quartic inflation, contrasting the cases of a positive and negative curvature of the inflaton-Higgs field-space manifold, corresponding respectively to $C_6=1$ (top) and $C_6=-1$ (bottom).
As the figure immediately shows, these operators drastically influence the cosmological fate of the Higgs.

As discussed in section~\ref{stochastic}, as inflation proceeds the dimension-six operator introduces two effects in the Higgs dynamics. For positive (respectively negative) curvature of the field-space manifold, the classical dynamics is modified by a stabilizing (resp. destabilizing) contribution
arising in the effective potential $\propto C_6 \epsilon h^2$. At the same time, the corresponding increase (resp. decrease) of the mass of the Higgs' fluctuations reduces (resp. enhances) the amplitude of the stochastic noise.
Both effects act in the same direction in the two cases. They tend to decrease the variance of the PDF for positive curvature, and to increase it for negative one. As a consequence, the overall (de)stabilizing effect is visible in figure \ref{survival}: for positive curvature of the field-space manifold (top right plot), 
the red region shrinks and the red region expands, while a negative curvature has the opposite effect (bottom right plot). Quantitatively, this implies that, for given values of $\xi$ and SM parameters, varying the details of Planck-suppressed couplings between the Higgs and the inlaton modifies the constraint on the Hubble scale by orders of magnitude, which is rather remarkable.

\begin{figure*}[t!]
	\hspace{-0.55cm}
	\begin{subfigure}[t]{0.5\textwidth}
		\centering
		\includegraphics[width=1\linewidth]{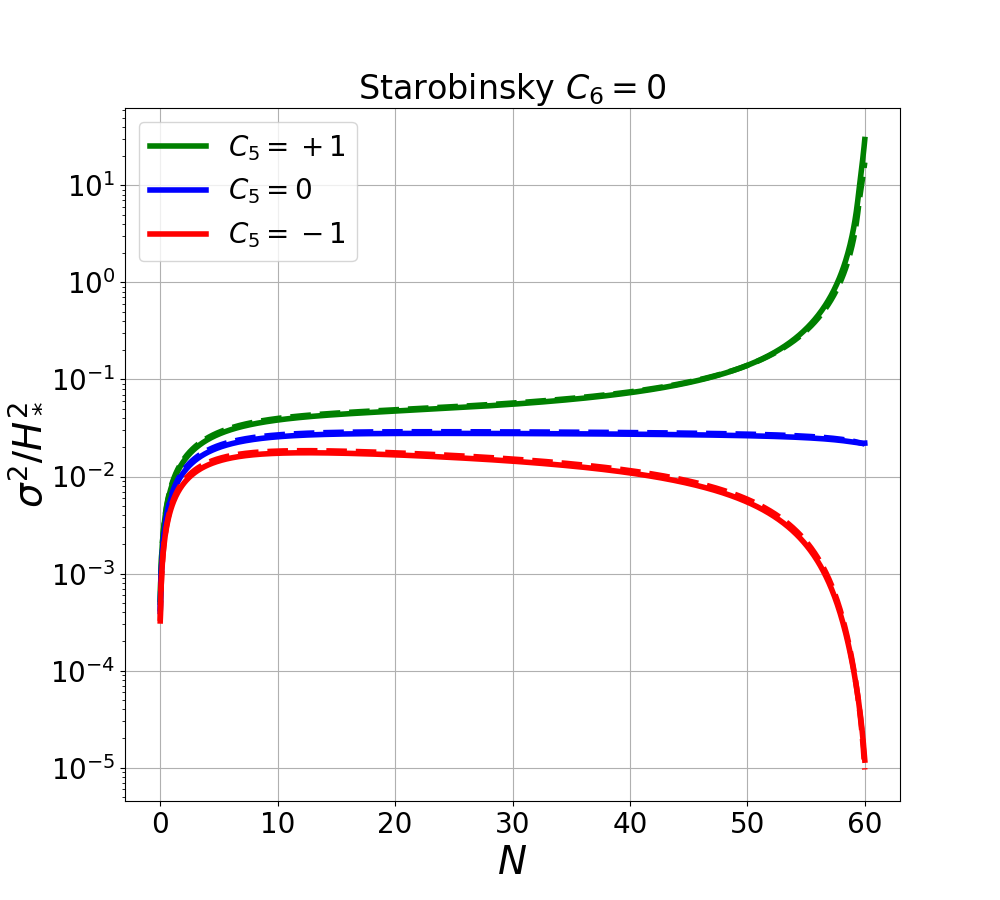}
		\caption{\textit{Evolution of the variance of the Higgs distribution (normalized to $H_\star$) for $\xi=-0.03$. For each case the full versus dashed lines represents the evolution determined by the conventional Fokker-Planck equation \eqref{langevin} versus the phase-space one \eqref{W} discussed in \ref{fullstoca}.}}
		\label{effect-b1}
	\end{subfigure}%
	~  \hspace{0.55cm}
	\begin{subfigure}[t]{0.5\textwidth}
		\centering
		\includegraphics[width=1\linewidth]{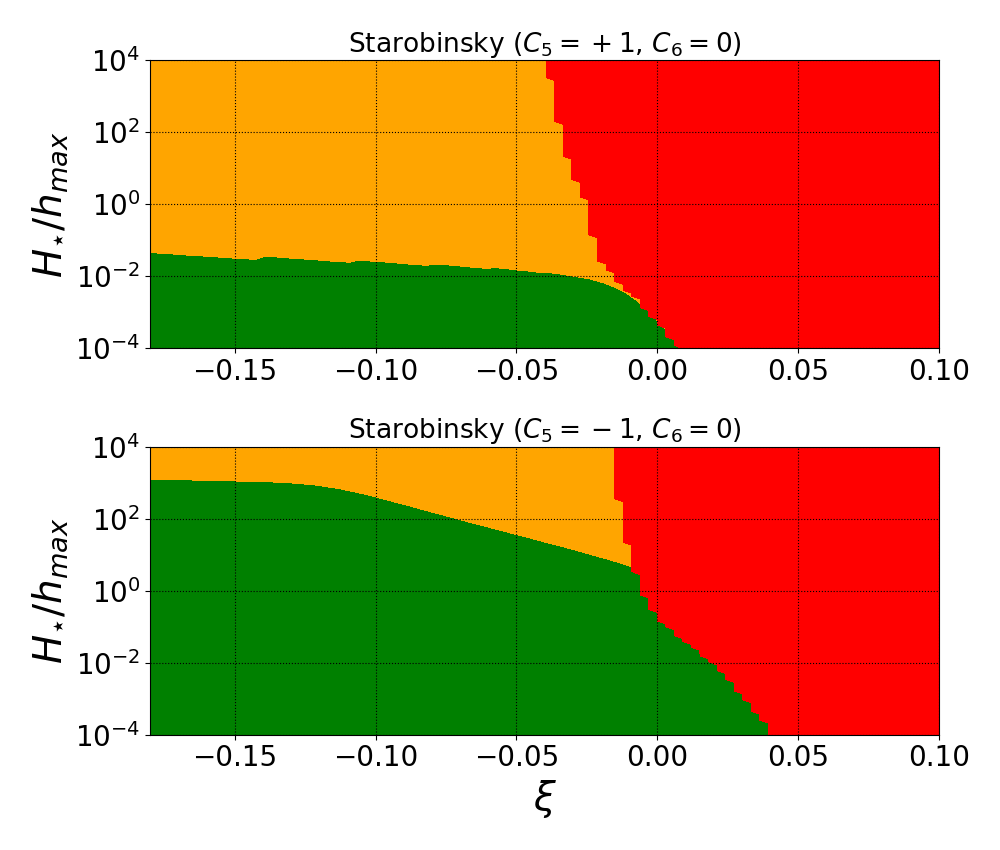}
		\caption{\textit{Like in figure \ref{survival}, green, orange and red regions correspond respectively to safe, potentially unsafe and excluded scenarios.}}	\label{survival-b1}
	\end{subfigure}
	\caption{\textit{Impact of the dimension-five operator in \eqref{kinetic-term} on the fate of the Higgs for Starobinsky inflation. }}
	\label{effect-b1-survival}
\end{figure*}

Let us also highlight the impact on our results caused by considering the finite time to fall into AdS.
To compute the fraction of patches in AdS, we take the maximum of $\mathcal{P}(|h|>h_{\mathrm{cl}})$ in the domain \eqref{settime}, which is cut at a given $e$-fold by requiring that $\hcl(N)$ has enough time to fall into AdS before inflation ends. For positive curvature, this maximum always occurs
before the time when there would not be enough $e$-folds left to fall into AdS. 
Hence including our cut does not affect the final results.
Conversely, for negative curvature, $\mathcal{P}(|h|>h_{\mathrm{cl}})$ keeps growing until the end of inflation, and it is therefore crucial to take into account this cut in order not to  significantly overestimate the effect of the derivative Planck-suppressed operator.\\[-0.3cm]

Finally, it is worth 
pointing out how the impact of derivative higher-order operators on the bounds is tight to the underlying background evolution. For instance, the dimension-six operator has tiny effects on Starobinsky-like evolutions (indeed fig.~\ref{survival1} would slightly change only for the case of negative curvature), for analogous reasons to those implying that plateau models lead to bounds close to the ones found in the de Sitter approximation,
i.e. the change in $\mathcal{P}(|h|>h_{\mathrm{cl}})$ does not happen early enough during inflation.
On the contrary, the dimension-five operator, as it gives a contribution to the effective potential $\propto C_5\sqrt{\epsilon}h^2$ (instead of $\propto\epsilon$), 
already influences significantly Starobinsky-like models, and has an even more dramatic impact on quartic inflation. For completeness, we illustrate explicitly the effect of the dimension-five operator on Starobinsky-like inflation in figure \ref{effect-b1-survival}. In figure \ref{effect-b1} we show the evolution of the variance for a given value of $\xi$ and Wilson coefficients $C_5=\{+1,0,-1\}$.
In comparison to figure \ref{derivative-operators-plot} (where deviations from the situation with no Planck-suppressed operator occur only in the last 10 $e$-folds for Starobinsky-like inflation), deviations occur earlier and also lead to a variance that is orders of magnitude different at the end of inflation. Indeed, the overall (de)stabilizing effects, represented in figure \ref{survival-b1} for $C_5=\pm 1$, are important throughout parameter space, with magnitudes similar to the effects that the dimension-6 operator has on quartic inflation (which is expected, as $\epsilon_{\mathrm{quartic}}\sim\sqrt{\epsilon_{\mathrm{Starobinsky}}}$).


\begin{figure*}[t!]
	\hspace{-0.55cm}
    \begin{subfigure}[t]{0.5\textwidth}
	\centering
	
	\includegraphics[width=1.0\linewidth]{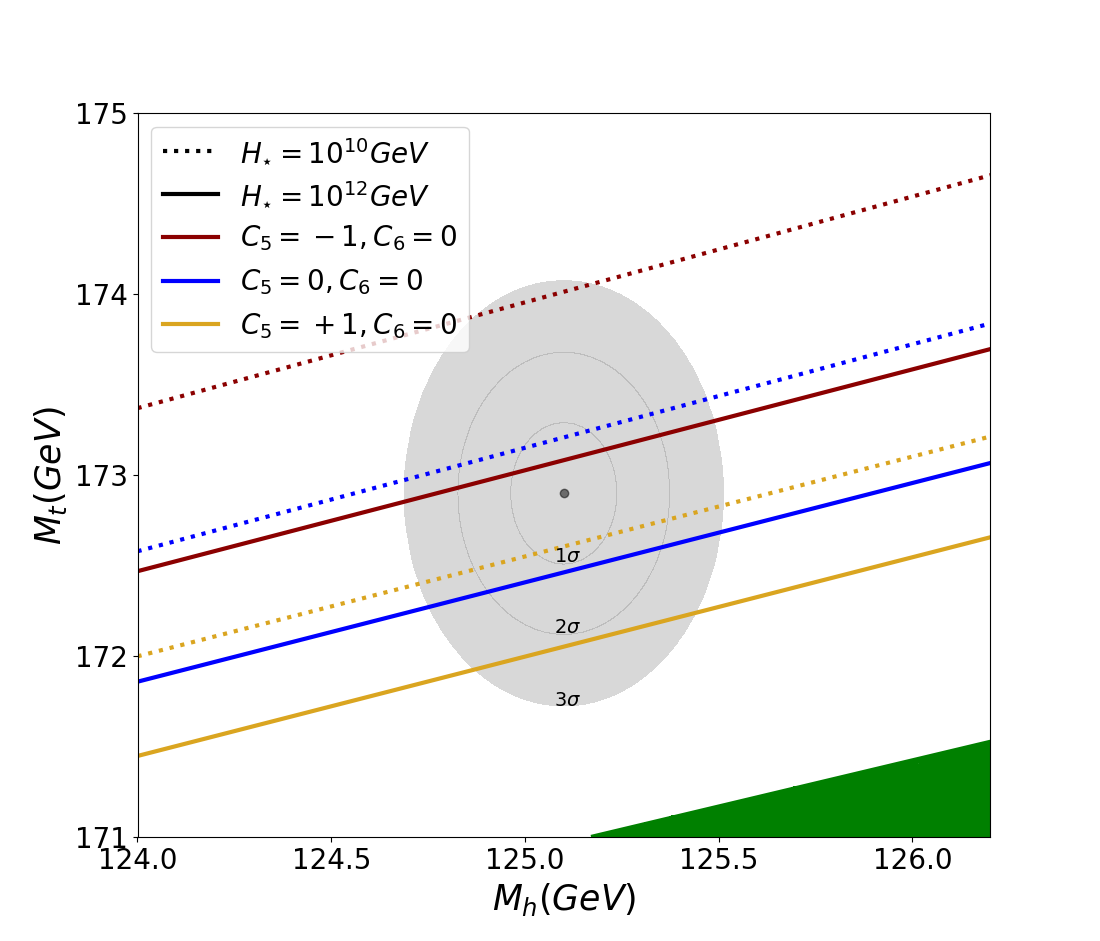}
	\caption{\textit{Effect of the dimension-5 operator on Starobinsky-like models.}}
	\label{mhmf1}
    \end{subfigure}%
    ~ \hspace{0.55cm}
    \begin{subfigure}[t]{0.5\textwidth}
	\centering
	\includegraphics[width=1.0\linewidth]{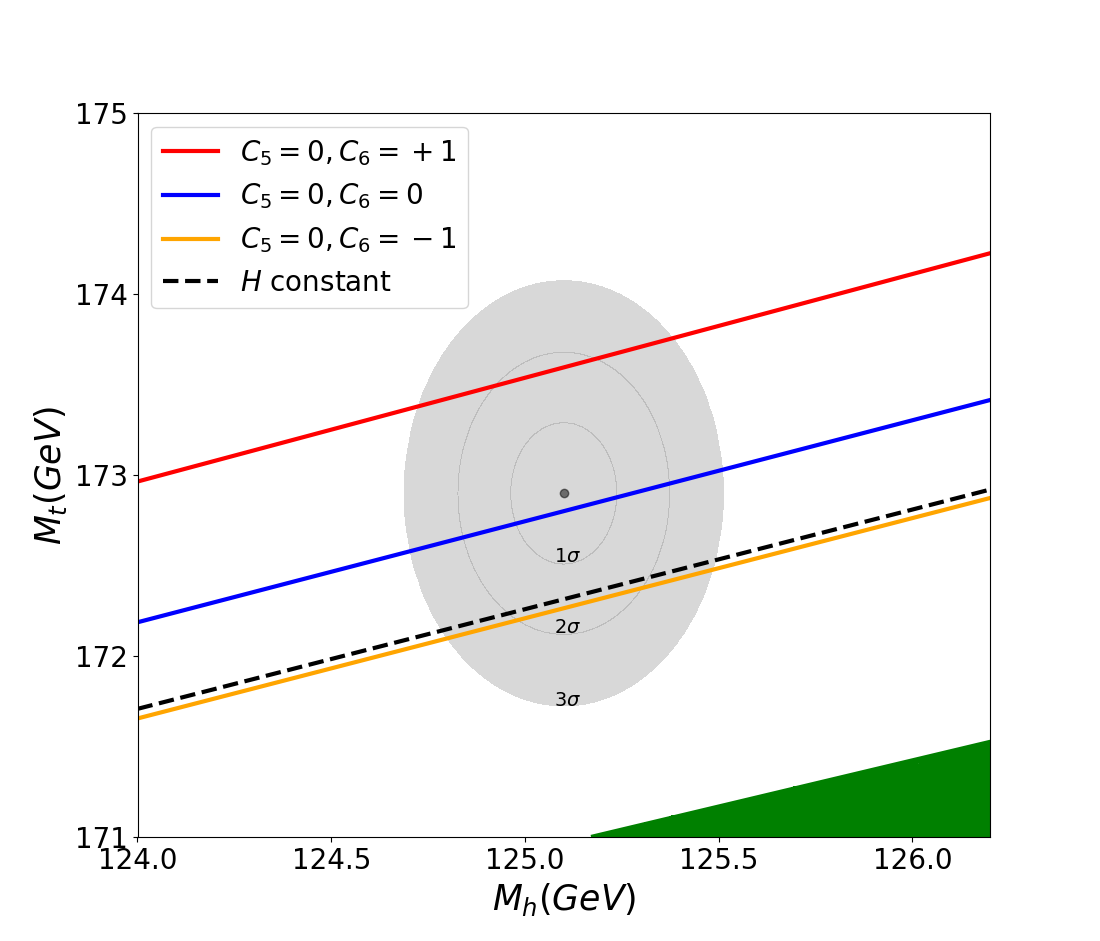}
	\caption{\textit{Effect of the dimension-6 operator on quartic-like models for $H=10^{12}$ $\mathrm{GeV}$.}}
	\label{mhmf}
    \end{subfigure}
    \caption{\textit{Bounds on the top/Higgs masses from the energy scale of inflation for $\xi=-0.05$, $\alpha_s=0.1181$ and $T_{\mathrm{RH}}\lesssim 10^4\, \mathrm{GeV}$ (equivalent to demanding no Hubble patch with $|h|>h_{\mathrm {max}}$ at the end of inflation), for different setups listed in the legend. Any given scenario marks a line separating the excluded region in parameter space (above) from the allowed one (below). In green, the stability region where the quartic Higgs coupling stays positive up to the Planck scale. The dashed black line on the right plot stays for the benchmark analysis done assuming $H=10^{12}\,\mathrm{GeV}=\mathrm{const}$ and $f=1$.}}
    \label{mtt}
\end{figure*}

\begin{figure*}[t!]
		\hspace{-0.55cm}
    \begin{subfigure}[t]{0.5\textwidth}
	\centering
	\includegraphics[width=1.1\linewidth]{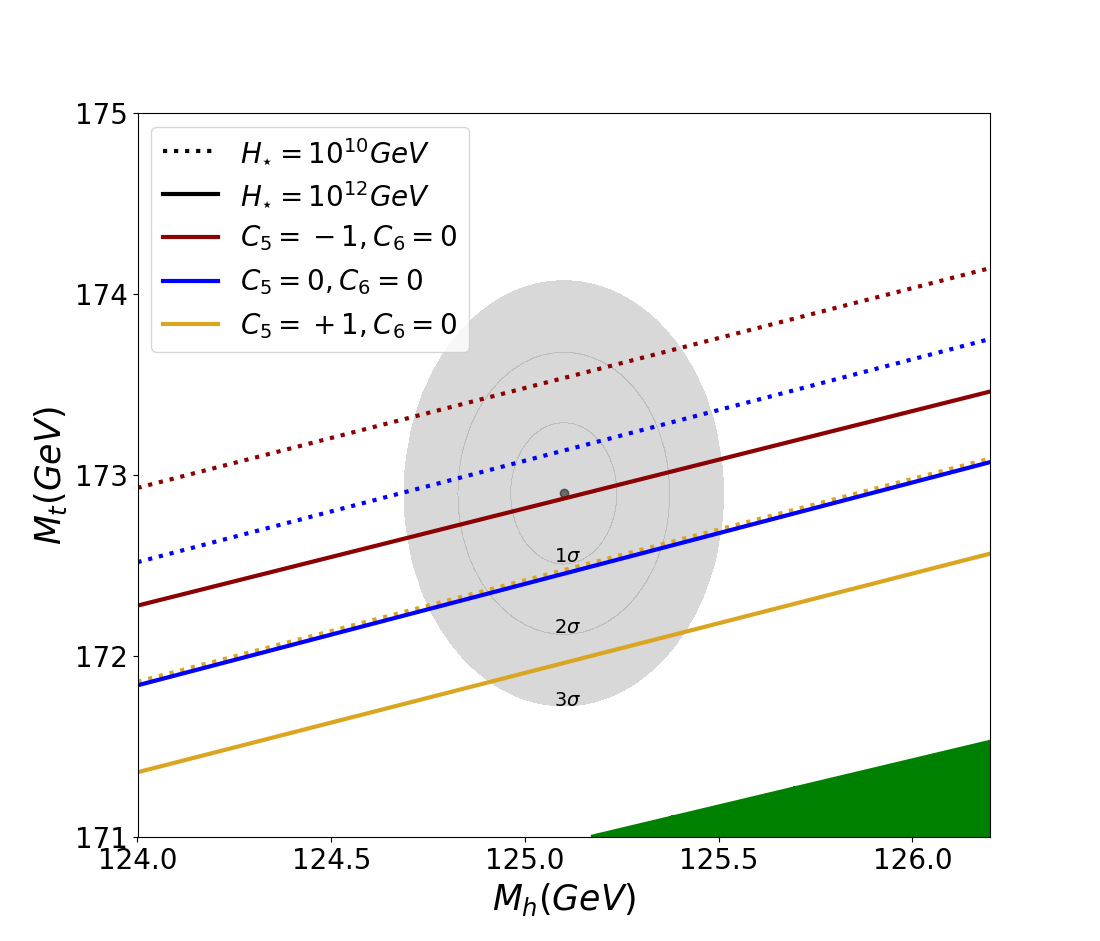}
	\caption{\textit{Effect of the dimension-5 operator on Starobinsky-like models.}}
	\label{mhaf1}
    \end{subfigure}%
    ~ 	\hspace{0.55cm}
    \begin{subfigure}[t]{0.5\textwidth}
	\centering
	\includegraphics[width=1.1\linewidth]{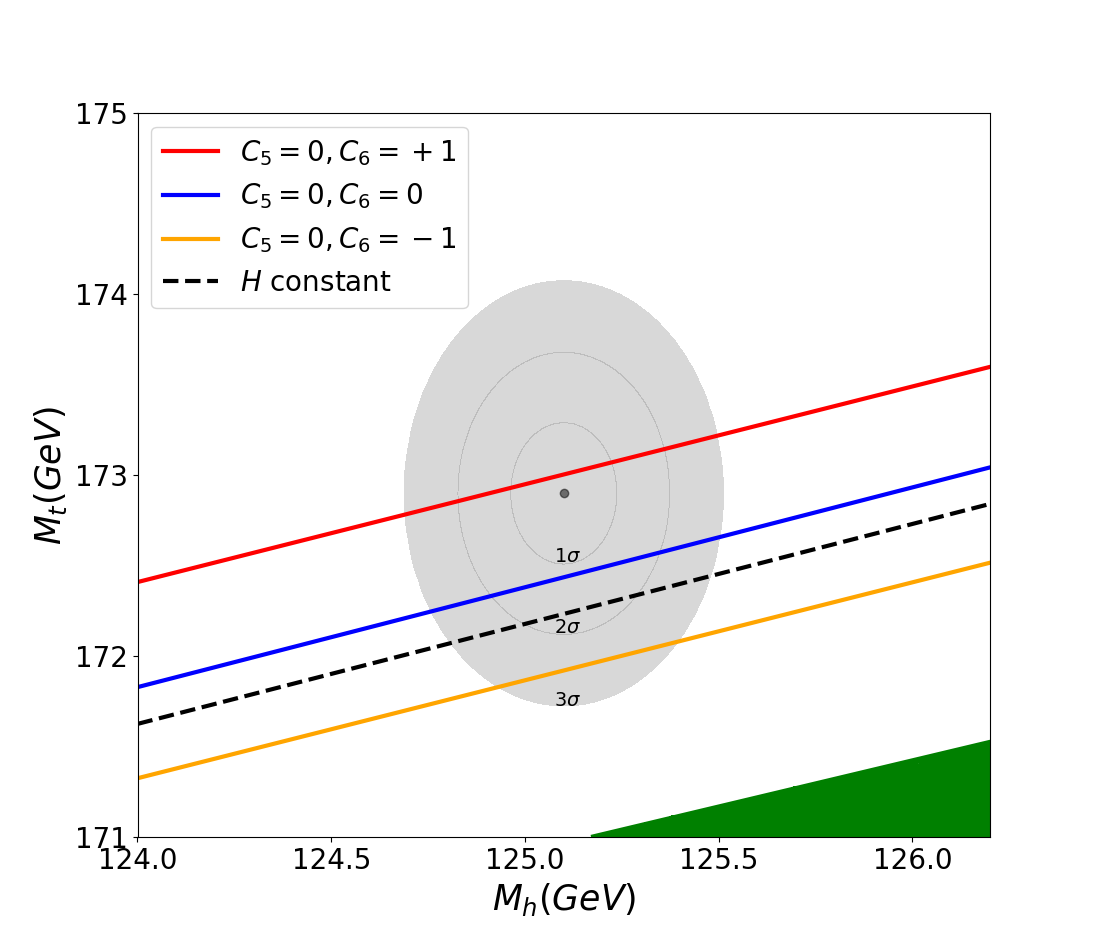}
	\caption{\textit{Effect of dimension-6 operator on quartic-like models for $H=10^{12}$ $\mathrm{GeV}$.}}
	\label{mhaf}
    \end{subfigure}
    \caption{\textit{Bounds on the top/Higgs masses from the energy scale of inflation for $\xi=-0.01$, $\alpha_s=0.1181$ and $T_{\mathrm{RH}}\gtrsim 10^{14}\, \mathrm{GeV}$(equivalent to demanding no Hubble patch in AdS at the end of inflation), for different setups listed in the legend. Any given scenario marks a line separating the excluded region in parameter space (above) from the allowed one (below). In green, the stability region where the quartic Higgs coupling stays positive up to the Planck scale. The dashed black line on the right plot stays for the benchmark analysis done assuming $H=10^{12}\,\mathrm{GeV}=\mathrm{const}$ and $f=1$.}}
    \label{mhh}
\end{figure*}

\subsection{Bounds on Standard Model parameters}\label{secbou}

The study of the Higgs instability during inflation can be equivalently applied to constrain the SM parameters within their experimental error bars. In fact, as already mentioned, the running of the Higgs quartic coupling is highly sensitive to the EW boundary conditions. 
Thus, it is instructive to look at the outcomes of our analysis from this different perspective.

The SM parameters over which our results are sensitive to are the top and Higgs masses $(M_t,M_h)$ and the strong coupling constant $\alpha_s$. The biggest uncertainty (both experimental and theoretical) comes from determining $M_t$ (see \cite{Hoang:2014oea,Nason:2017cxd} for recent discussions on the subject). For illustrative purposes, in order to compare our results with experimental data, we fix $\alpha_s$ to its central value $\alpha_s=0.1181$ \cite{Tanabashi:2018oca} and we vary $(M_t,M_h)$ within the five-sigma boundaries from their best current estimate:
$M_h=125.10\pm0.14\,\mathrm{GeV}$ and $M_t=172.9\pm0.4\,\mathrm{GeV}$ \cite{Tanabashi:2018oca}. For $M_t$ in particular we take the quoted direct measurements value from the PDG \cite{Tanabashi:2018oca}.
For each value of $\xi$, a given $H_\star$, hopefully given in the future by a detection of primordial B-modes, marks a line dividing the $(M_t,M_h)$ plane in two regions: below, the allowed (safe) region in parameter space (no patches in which the Higgs has fluctuated above the potential barrier), above the (potentially) unsafe region in which dangerous patches have formed (meaning regions in which $h>h_{\mathrm{max}}$ in figure \ref{mtt}, or AdS regions in figure \ref{mhh}). 
In the same manner as in the previous section,
for a few parameters of interest, we study how these bounds change once the various 
effects considered in this work are taken into account.\\

In figure \ref{mtt} we fix $\xi=-0.05$ and consider  $\mathcal{F}_{|h|>h_{\mathrm {max}}}$, so that each setup provides a line separating the safe region below
from the potentially unsafe one (above), equivalently to the boundary between the green and orange regions in figure~\ref{survival}.
Two scales of inflation are used in figure \ref{mhmf1}, namely $H_{\star}=10^{10}\, \mathrm{GeV}$ and $H_{\star}=10^{12}\, \mathrm{GeV}$, for Starobinsky-like inflationary evolutions. For each of them, we considered the dimension-five operator (alone) by varying $C_5 =0,\pm 1$. In the right figure \ref{mhmf}, we consider the dimension-6 operator alone, $C_6=0,\pm 1$, this time for quartic inflation and for the value $H_*=10^{12}\,\mathrm{GeV}$.
This energy scale corresponds to a tensor-to-scalar ratio of order $\sim 10^{-4}$, the lowest one that can be observationally probed in the near future \cite{Abazajian:2016yjj}.

Unsurprisingly, increasing $H_{\star}$ always shrinks the allowed region. 
This has been shown previously in \cite{Hook:2014uia, Kearney:2015vba,Franciolini:2018ebs} and is easy to understand; larger $H_\star$ imply larger stochastic kicks. Thus, under the same conditions, it is more likely that the Higgs ends up beyond the potential barrier.
More interestingly, depending on the sign of the Wilson coefficient, for plateau-like models the dimension-5 operator has an important
stabilizing or destabilizing effect, similar to the one the dimension-6 operator has on quartic-like models (even if the latter is larger) 
for the reason mentioned in the previous section. Indeed, one can see in figure \ref{mhmf1} that a change of $H_\star$ by two orders of magnitude can be otherwise mimicked by simply considering the effects of a Planck-suppressed operator (see for example the solid and dashed blue lines versus the solid blue and solid brown). Eventually, in figure~\ref{mhmf}, the dashed black line marks the boundary for the benchmark study, in which the time dependence of the background and the deviation from massless noise ($f\neq1$) are not taken into account. The appreciable difference between the dashed black and the blue solid line thus highlights the importance to include these effects in the analysis.\\[-0.3cm]

In figure \ref{mhh} we consider $\mathcal{F}_{\mathrm{AdS}}$, the fraction of patches already in AdS at the end of inflation, with the same setup as in figure \ref{mtt} but here for $\xi=-0.01$. The various lines, equivalent to the boundary of the red regions in figure \ref{survival}, split the parameter space between the allowed region (below) and the excluded one (above), which cannot be rescued by any post-inflationary dynamics.
Given the exponential sensitivity of the bound \eqref{prob-result} to models' parameters, that we understood analytically in Sec.~\ref{secmod1},
for larger values of $\xi$, the shift of the various lines can be as pronounced as to exit the five-sigma contours, meaning a complete rescuing effect. In the current example of $\xi=-0.01$ (chosen for illustrative purposes), one can already see how easily the effects studied in this work can alleviate possible tensions between the central values of measured SM parameters and typical expected values for the energy scale of inflation.

\section{Conclusions}\label{seccon}

We revisited the important question of the stability of the Higgs vacuum during inflation, by taking into account features of realistic models that have been hitherto overlooked: the unavoidable time-dependence of the Hubble scale during inflation, and the generic presence of derivative operators coupling the Higgs and the inflaton. A motivation for looking at the latter aspect is the well known fact that higher-order operators suppressed by a high energy scale can have a critical impact on effective masses of scalar fields during inflation, as exemplified by the eta-problem and the geometrical destabilization of inflation.

We studied these aspects in a simple but rather generic manner. 
We considered different inflationary backgrounds and enlarged usual setups by considering two-derivative higher-order operators that are inflaton shift-symmetric, keeping track of the effects of dangerous irrelevant operators. We focused for simplicity on operators suppressed by the Planck scale, as we demonstrate that even these ones have significant consequences.
We showed that one can initially neglect the backreaction of the Higgs on the inflaton, and consider that the former undergoes a stochastic motion subject to a time-dependent effective potential. This comprises not only to the SM potential and quadratic potential induced by the non-minimal coupling of the Higgs, like in previous studies. It also has two additional quadratic contributions generated by specific dimension-5 and dimension-6 operators. The corresponding induced masses squared, in Hubble units, can assume any sign and are proportional respectively to $\sqrt{\epsilon}$ and $\epsilon=-\dot{H}/H^2$, the usual slow-roll parameter, as a consequence of their kinetic origin. Therefore, their quantitative impact depends on the specific evolution of the Hubble scale during inflation, and is inevitably tied to the other aspect that distinguishes our work from previous ones, i.e. considering the time-dependence of the inflationary background. We stress that despite the apparent smallness of these mass terms, they can have a crucial impact on the cosmological fate of the Higgs vacuum. \\[-0.3cm]

We considered the Fokker-Planck equation that governs the evolution of the distribution of Higgs' values in Hubble-sized regions. 
We showed explicitly that the effects caused by the time dependence of the background, Planck-suppressed derivative operators, and the stochastic noise of light fields differing from the one of exactly massless ones, have important consequences for the distribution of Higgs values, and hence for the fate of the Higgs. 

Previous works showed that not a single Hubble patch in our observable universe at the beginning of the radiation era should be such that the Higgs reached sufficiently large values as to form an AdS patch, i.e. a crunching region surrounded by a causally disconnected one of negative energy density. However, not all patches in which the Higgs has fluctuated above the potential barrier share this fate. Depending on the reheating temperature, thermal corrections to the Higgs potential can go from rescuing regions (which are not yet in AdS) with arbitrarily large Higgs values, to rescuing none. Therefore, we used two different criteria, corresponding to these two extreme situations, to qualify each model either as excluded, allowed, or potentially unsafe. By doing so, and owing to the inherent time-dependence of our effective potential, we had to pay attention to the fact that patches already in AdS cannot be rescued, as well as to the finite time it takes for them to form when the Higgs backreaction cannot be neglected anymore, resulting in a new procedure explained in section \ref{section3}.

In our numerical analysis, we considered two different inflationary backgrounds, corresponding to Starobinsky and quartic inflation, meant as representative of models with respectively negligible and appreciable time dependence of the Hubble scale in the bulk of the inflationary phase. We also varied the Wilson coefficients of the dangerous dimension-5 and -6 operators, the value of the non-minimal coupling and the overall Hubble scale. As for the purely Standard Model sector, we varied the Higgs and top masses measured at the electroweak scale within their experimental error bars.
An obviously important parameter for the fate of the Higgs is the ratio $H_\star/\hmax$ between the Hubble scale, setting the overall amplitude of stochastic kicks, and the location of the potential barrier. This shows that results can be seen from two complementary perspectives: as bounds on the energy scale of inflation, or as bounds on SM parameters governing the location of the potential barrier. We adopted the two viewpoints and summarized our main numerical results in figures \ref{survival} and \ref{effect-b1-survival} (first perspective) and \ref{mtt} and \ref{mhh} (second perspective), contrasting them with previous similar figures in the literature that do not take into account aspects developed in this work.

Besides the precise understanding that we gained of how the different effects we took into account affect the fate of the Higgs, we can draw two general lessons from these results. The first is that, for given SM parameters and scale of inflation, different time-dependence of the Hubble scale and Planck-suppressed couplings between the Higgs and the inflaton, which may appear as unimportant details, can lead to radically different outcomes for the fate of the instability, turning an allowed model into an excluded one and vice-versa.
The second related lesson is that, with the existence of a degeneracy between values of the Hubble scale separated by several orders of magnitude on one side, and effects coming from Planck-suppressed couplings on the other side, it appears unlikely that a future detection of primordial gravitational waves would, on its own, enable one to constrain efficiently SM parameters.\\[-0.3cm]

Our work offers natural avenues for future studies in different directions. For the first time in the study of the Higgs vacuum instability, we have taken into account the time-dependence of the background and the amplitude of the noise differing from the one of exactly massless fields. We did so in simple motivated manners, but given the important impact of these aspects, it would be useful to develop a more thorough theoretical understanding of them, and more generally of the theoretical uncertainties of the stochastic formalism (see discussions in section \ref{stochastic}). It would also be worthwile to go beyond the Gaussian approximation for the PDF in our time-dependent background (see \cite{Kearney:2015vba,East:2016anr,Jain:2019wxo} for such studies in de Sitter). For instance, challenging as it is, one can envisage to solve numerically the Fokker-Planck equation with the quartic potential taken into account, and with suitable time-dependent boundary conditions that incorporate the transmutation of inflationary patches into AdS ones at large Higgs values. Eventually, it would be interesting to revisit the generation of primordial black holes in the Standard Model \cite{Espinosa:2017sgp} by taking into account the aspects developed in this work. In particular, the various setups that we studied lead to different times to form an AdS region, which could alleviate the fine-tuning problem behind this proposal
\cite{Gross:2018ivp,Espinosa:2018euj}. 
\newpage
\subsection*{Acknowledgments}

We are grateful to V. Branchina, P. Carrilho, J.R. Espinosa, G. Franciolini, T. Markkanen, D. Mulryne, L. Pinol, M. Postma, D. Racco, A. Riotto, A. Shkerin, T. Terada, V. Vennin, L. Witkowski for interesting and helpful discussions, as well as to the referee who helped improving our paper. J.F, S.RP and J.W.R are supported by the European Research Council under the European Union's Horizon 2020 research and innovation programme (grant agreement No 758792, project GEODESI).
	
\appendix
\section{Time to fall into AdS}
\label{time-ads}

\begin{figure*}[t!]
\hspace{-0.55cm} 
 \begin{subfigure}[t]{0.5\textwidth}
    \centering
	\includegraphics[width=1.1\linewidth]{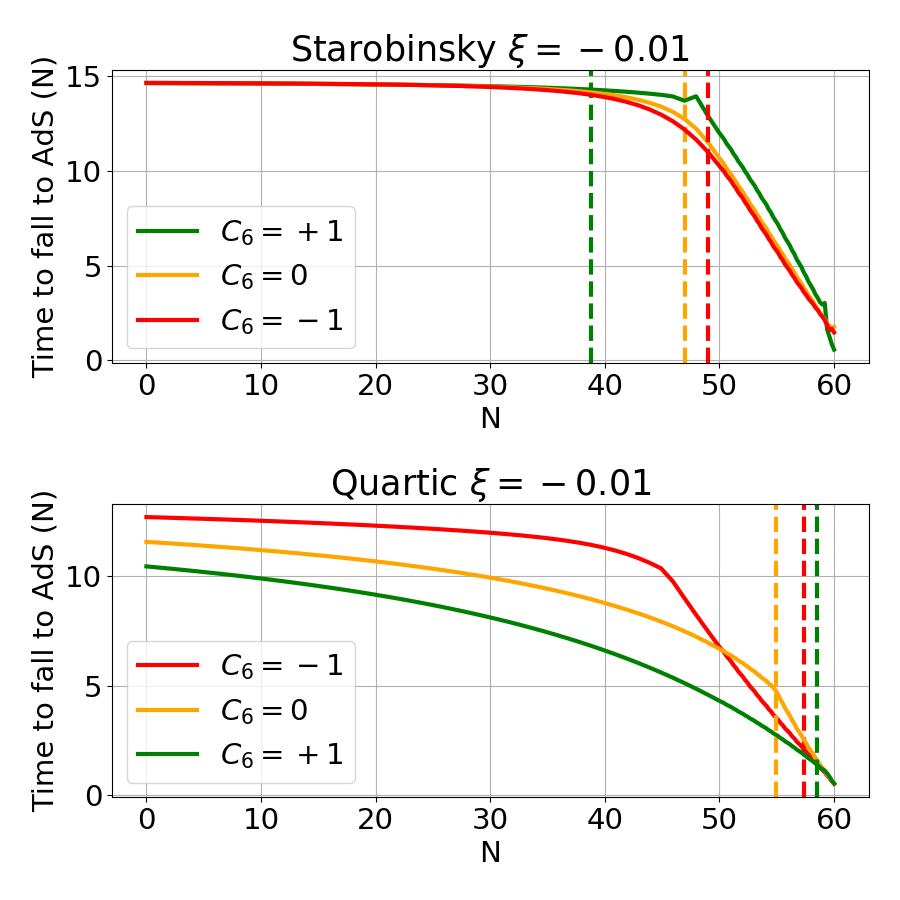}
	\caption{\textit{Time needed to form an AdS region versus the number of $e$-folds. For each scenario, the dashed vertical line indicates the time 
		relevant to compute the fraction of patches in AdS at the end of inflation (see Eq.~\eqref{prob-result}).}}
	\label{mhaf2}
\end{subfigure}%
    ~\hspace{0.55cm}
\begin{subfigure}[t]{0.5\textwidth}
    \centering
	\includegraphics[width=1.1\linewidth]{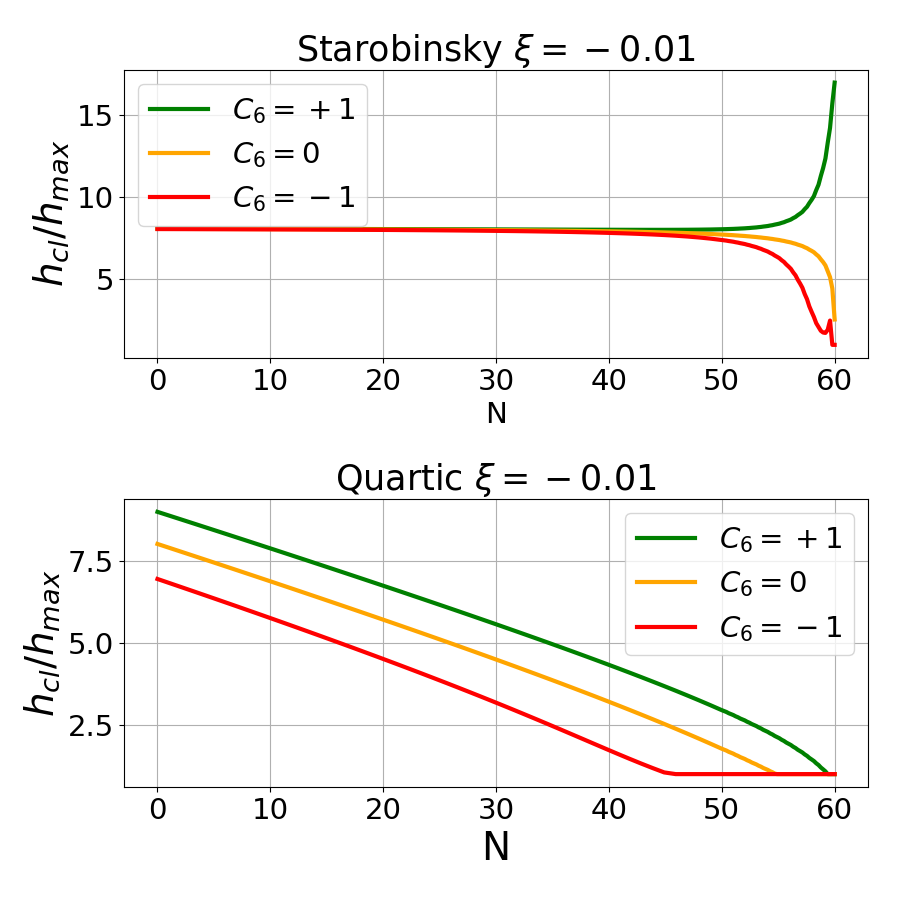}
	\caption{\textit{Evolution of $\hcl$ normalized to $\hmax$, in time-dependent backgrounds.  $\hcl$, defined in Eq. \eqref{had}, is the point beyond which the classical dynamics dominates over the stochastic one, and the Higgs starts to fall towards the true vacuum.}}
	\label{vev}
    \end{subfigure}
    \caption{\textit{Time to form an AdS region starting from the ``point of classicality" $\hcl$.}}
	\label{vev2}
\end{figure*}

As explained in Sec.~\ref{matching}, 
it is particularly important in a time-dependent setup to take into account, 
at any step of the evolution, the finite time to form an AdS region starting from $h=\hcl(N)$, the point after which stochastic kicks are not relevant anymore. Around $\hcl$, the energy is still dominated by the inflaton sector and hence, it is a good approximation to consider the Higgs as a spectator field until that point. However, the Higgs backreaction on the background cannot be neglected anymore when the Higgs falls towards the true vacuum. Thus, from $\hcl$ onwards we evolve the full two-field system classically. We consider the formation of an AdS patch unavoidable when the Hubble scale in the Einstein frame becomes negative. 
This corresponds to the onset of the development of a shrinking region, which has been shown in \cite{East:2016anr} to lead to an AdS region (see the introduction).\footnote{We use as a criterion $H_E<0$ in the Einstein frame because the numerical simulations in \cite{East:2016anr} 
were performed without the non-minimal coupling between the Higgs and the Ricci scalar, i.e. in the Einstein frame.} 
As an aside, this prescription is even more optimistic than the ``most optimistic possibility" in Ref. \cite{Gross:2018ivp}, i.e. $V_{\mathrm{eff}}(h)+V(\phi)=0$.

In figure ~\ref{vev2} we plot the time required to fall into AdS from $\hcl$ onwards (left), and the time evolution of $\hcl$ itself (right), for a given $\xi$, two different background evolutions and different choices of Wilson coefficient for the six-dimensional operator.
For simplicity and given the illustrative scope of this appendix, we restrict to the approximate form of the potential \eqref{appro2} for the central values of SM parameters.\footnote{The difference between the analytical approximation and the full NNLO numerical potential is that the approximate one has a steeper drop-off after $\hcl$. Classically evolving from $\hcl$ to the point where the energy density becomes negative means the Higgs field
gathers more kinetic energy and arrives at that point earlier (a shift of about 3 $e$-folds).}
The dashed vertical lines correspond to the time  at which the maximum of $\mathcal{P}(|h|>h_{\rm cl}(N),N)$ is taken (at $N=\Nm$, see Eq.~\eqref{prob-result}) in the domain $\hcl \in \mathrm{AdS}$ defined in \eqref{settime}, i.e. where there is still enough time left before the end of inflation to form an AdS patch starting 
from $\hcl(N)$.

As an example, consider Starobinsky inflation with $C_6=1$ (the green line in the top left panel). In that case, the variance grows and subsequently decreases, and the time $\Nm$ in \eqref{prob-result} coincides with the time of the maximum of $\mathcal{P}(|h|>h_{\rm cl}(N),N)$ during the whole inflationary evolution: 
21.2 $e$-folds before the end of inflation, sufficiently long enough for the required 14.7 $e$-folds to fall into AdS.
In contrast, for $C_6=-1$ (the red line in the top left panel), the destabilization driven by the negative curvature of the Higgs-inflaton manifold causes the variance to grow until the end of inflation, as well as $\mathcal{P}(|h|>h_{\rm cl}(N),N)$. In this case, $\Nm$ does not coincide with the time of the maximum of $\mathcal{P}(|h|>h_{\rm cl}(N),N)$ during the whole inflationary evolution; it occurs 11 $e$-folds before the end of inflation, when there is still enough time to fall into AdS. This underlines the importance of considering the finite time to fall into AdS to properly estimate the fraction of AdS patches at the end of inflation. 

In figure \ref{vev} we plot the values of $\hcl$ as a function of time. 
This gives the reader an indication of the initial field value of the Higgs at the start of the classical evolution. As for the initial velocity of the Higgs, we set $\dot{h}=0$ (even if it is almost instantaneously attracted towards the slow-roll velocity).
The initial conditions for the field $\phi$ at the start of the classical evolution are the same as the ones it would have had at that time during inflation with the Higgs as a spectator field. 
Something important to note: 
one might naively expect that the smaller $\hcl$, the larger the time to fall into AdS. However, this deceptive intuition does not take into account the full two-field evolution. In this framework, the drop in overall energy density due to the inflationary field rolling towards the end of inflation dominates over the decrease of $\hcl$. 
This is rather remarkable, as it can change the time to fall into AdS by a substantial amount, as can be seen by comparing the two models in figure \ref{mhaf2}.

\bibliographystyle{utphys}
\bibliography{Biblio2019}

\end{document}